\documentclass[article]{aa}
\usepackage{multirow}
\usepackage{amsmath}
\usepackage{graphicx,subfig} 
\usepackage{lscape}
\usepackage{longtable}
\usepackage{txfonts}
\usepackage{booktabs}
\usepackage{soul}
\usepackage{rotating}
\usepackage{natbib,twoopt}
\usepackage{xcolor}
\usepackage{siunitx}
\usepackage[dvipsnames]{xcolor}

\makeatletter
\renewcommand*\aa@pageof{, page \thepage{} of \pageref*{LastPage}}
\usepackage{hyperref}
\hypersetup{colorlinks=true,allcolors=[rgb]{0,0,0.8}}

\renewcommand{\arraystretch}{1.2}
\newcommand{\rosi}{eROSITA\xspace}

\newcommand{\erass}{eRASS1\xspace}
\newcommand{\extlike}{$\mathcal{L}_{\rm ext}$}

\begin{document}

\title{The SRG/eROSITA all-sky survey}

\subtitle{X-ray scaling relations of galaxy groups and clusters\\in the western Galactic hemisphere}

\author{M.~E.~Ramos-Ceja\inst{1,}\thanks{These authors contributed equally to this work.}, L.~Fiorino\inst{1,}\footnotemark[1], E.~Bulbul\inst{1,2}, V.~Ghirardini\inst{3}, N.~Clerc\inst{4}, A.~Liu\inst{5}, J.~S.~Sanders\inst{1}, Y.~E.~Bahar \inst{3} J.~Dietl\inst{6}, M.~Kluge \inst{1}, F.~Pacaud\inst{6}
\and E.~Artis \inst{1}
\and F.~Balzer \inst{1}
\and J.~Comparat \inst{1}
\and Z.~Ding \inst{1}
\and N.~Malavasi \inst{1}
\and A.~Merloni \inst{1}
\and T.~Mistele \inst{1}
\and K.~Nandra \inst{1}
\and R.~Seppi \inst{7}
\and S.~Zelmer \inst{1}
\and X.~Zhang \inst{1}
}
\institute{
Max Planck Institute for extraterrestrial Physics, Giessenbachstrasse 1, 85748 Garching, Germany \\
\email{mramos@mpe.mpg.de}
\and
Faculty of Physics, Fakult\"at f\"ur Physik, Ludwig-Maximilians-Universit\"at, Scheinerstr. 1, 81679 M\"unchen, Germany
\and
INAF, Osservatorio di Astrofisica e Scienza dello Spazio, Via Piero
Gobetti 93/3, 40129, Bologna, Italy
\and
IRAP, Universit{\'e} de Toulouse, CNRS, UPS, CNES, Toulouse, France
\and
Institute for Frontiers in Astronomy and Astrophysics, Beijing Normal University, Beijing 102206, China
\and
Argelander-Institut f{\"u}r Astronomie (AIfA), Universit{\"a}t Bonn, Auf dem H{\"u}gel 71, 53121 Bonn, Germany
\and
Department of Astronomy, University of Geneva, Ch. d’Ecogia 16, CH-1290 Versoix, Switzerland
}

\titlerunning{X-ray scaling relations of the eRASS1 galaxy groups and clusters}
\authorrunning{Ramos-Ceja, M. E., et al.}

\abstract{
The soft X-ray telescope on board the Spectrum-Roentgen-Gamma (SRG) mission, extended ROentgen Survey with an Imaging Telescope Array (eROSITA), has produced the largest sample to date of galaxy groups and clusters detected via their intracluster and intragroup medium emission. Scaling relations between the intrinsic properties of these systems provide valuable insight into their formation and evolution. In this work, we investigated the scaling relations between key physical properties, such as soft band X-ray luminosity, temperature, gas mass, and the low-scatter mass proxy $Y_{\rm X}$, for the galaxy groups and clusters detected in the first eROSITA All-Sky Survey (eRASS1). Our analysis fully accounts for selection effects and the redshift evolution of the observable distributions. We constructed a high-purity sample of $3061$ galaxy groups and clusters spanning the redshift range of $0.05<z<1.07$ and the mass range of $1.1\times10^{13}<M_{500}/$M$_{\odot}<1.6\times10^{15}$. This represents the largest sample to date used in scaling-relation analysis. The selection function, derived from state-of-the-art simulations of the eROSITA sky, is rigorously incorporated into our modeling. We report best-fit parameters -- normalization, slope, redshift evolution, and intrinsic scatter -- for a set of scaling relations: $L_{\mathrm{X}}-T$, $L_{\mathrm{X}}-M_{\rm gas}$, $L_{\mathrm{X}}-Y_{\rm X}$, as well as the $M_{\rm gas}-T$ relation. Our best-fit models indicate that the slopes of the scaling relations significantly deviate from self-similar expectations, while the redshift evolution remains consistent with the self-similar model. The fits exhibit small statistical uncertainties, likely owing to the large sample size. Our results are in good agreement with previous observational studies that account for selection effects. For clusters with $kT>3$~keV, our inferred $L_{\rm X}$--$T$ relation is consistent with simulation models that include strong or stronger AGN feedback.
}


\maketitle


%
\section{Introduction}
\label{sec:intro}
Galaxy clusters and groups retain a substantial fraction of baryons within their potential wells. They originate at the highest peaks of the cosmic density field, where matter departs from the Hubble flow and collapses under gravity into progressively larger haloes. In the simplest framework, clusters and groups result from scale-free gravitational collapse of the highest peaks in the early Universe \citep[e.g.,][]{Kravtsov2012}. This scenario implies that haloes of different masses and redshifts are scaled versions of one another, an assumption known as the self-similar model \citep{Kaiser1986}. Within this framework, observable cluster properties are expected to follow simple power-law scaling relations with total mass.

Scaling laws, encompassing both baryonic and nonbaryonic components, have long served as probes of the underlying cosmological model \citep[e.g.,][]{Reiprich2002,Vikhlinin2009,Mantz2010,Allen2011,Reichardt2013,Bulbul2019,Pratt2019,Ghirardini2024}. Besides being essential in cosmology, scaling relations provide constraints on the thermodynamic evolution of baryons in the intracluster and intragroup medium (ICM and IGrM) and enable us to quantify the impact of nongravitational processes \citep[e.g.,][]{Ettori2004,Maughan2007,Pratt2009,Sun2009,Vikhlinin2009,Mantz2010,Eckmiller2011,Kettula2015,Lovisari2015,Ghirardini2021,Bahar2022}. Departures from self-similarity arise from mechanisms that alter the ICM and IGrM gas distribution, such as feedback \citep[e.g.,][]{McNamara2007,Fabian2012,Gitti2012, Barnes2017}, bulk and turbulent motions \citep[e.g.,][]{Churazov2004,Zhuravleva2014}, and galactic winds \citep[e.g.,][]{Menci2000,Tozzi2001}. As a result, scaling relations are widely studied to measure these deviations and to identify low-scatter mass proxies for cosmological applications \citep{Kravtsov2006}. For instance, luminosity–mass relations in galaxy groups have been used to test the effects of active galactic nuclei (AGN) feedback \citep[e.g.,][]{Sun2009, Eckmiller2011, Lovisari2015, Bahar2022, Bahar2024}, while relations involving other observables, such as luminosity with the Compton-$y$ parameter or optical richness, are central to multiwavelength studies of clusters \citep[e.g.,][]{Zhang2011, Planck2013, IderChitham2020, Ramos-Ceja2022, Kluge2024}.

The soft X-ray telescope extended ROentgen Survey with an Imaging Telescope Array \citep[eROSITA,][]{Predehl2021}, aboard the Spectrum-Roentgen-Gamma (SRG) mission \citep{Sunyaev2021} was launched on July 13, 2019. With its large collecting area (1365 cm$^2$ at $1$~keV) and moderate angular resolution ($\sim30^{\prime\prime}$ at $1.49$~keV), eROSITA provides an unprecedented view of the X-ray sky. Its primary goal is to build the largest, well-characterized cluster samples, enabling cosmological constraints from cluster abundances that are complementary to cosmic microwave background (CMB) probes when combined with weak-lensing mass calibration \citep{Merloni2012}.

In this work, we used the largest ICM- and IGrM-selected sample of galaxy clusters and groups from the first eROSITA All-Sky Survey (eRASS1) to date to study X-ray scaling relations among key observables. The sample spans three orders of magnitude in mass and four in luminosity \citep{Bulbul2024}. With this dataset, we constrained the influence of nonthermal processes on intrinsic cluster properties and established links between galaxy groups, transitional systems, and massive clusters, fully exploiting the statistical power of the eROSITA catalogs \citep{Kluge2024, Merloni2024}. Previous analyses based on the smaller-area eROSITA Final Equatorial-Depth Survey (eFEDS) have provided valuable scaling relations but were limited by sample size \citep{Liu2022, Bahar2022, Chiu2022}. For this study, we extended those studies by a factor of two in sample size and probed haloes down to $1\times10^{13}$~M$_{\odot}$. In particular, we investigated the scaling relations of eRASS1 clusters and groups between soft band X-ray luminosity, temperature, gas mass, and the low-scatter mass proxy $Y_{\rm X}$.

This paper is organized as follows. In Section~\ref{sec:sample_analysis}, we describe the group and cluster sample selection, the data reduction, and analysis. In Section~\ref{sec:scaling_modeling}, we present the scaling relation fitting method and the selection function used for this work. X-ray scaling relations between the X-ray properties of the selected clusters are given in Section~\ref{sec:results}. We discuss our results in Section~\ref{sec:discussion}, and provide a summary of the presented work in Section~\ref{sec:summary}.

Throughout this paper, we use a $\Lambda$ Cold Dark Matter (CDM) cosmology, characterized by parameters from \cite{Planck2016}, namely $\Omega_\mathrm{m}=0.3089$, $\sigma_8 = 0.8147$, $H_0 = 67.74\ \mathrm{km \ s^{-1}\ Mpc^{-1}}$, and $n_\mathrm{s}=0.9667$. Quoted error bars correspond to a 1$\sigma$ confidence level unless noted otherwise.


%
\section{Sample selection and data analysis}
\label{sec:sample_analysis}

In this section, we present the selected cluster sample and the X-ray data analysis to extract its X-ray properties. The sample of sources and their X-ray properties are based on our previously published catalogs. We provide critical points on the sample selection and eROSITA X-ray analysis. 

\subsection{eROSITA galaxy clusters and groups sample}
\label{subsec:sample_selection}

We based our analysis on the eROSITA primary galaxy cluster and group catalog presented in \citet{Bulbul2024} and \citet{Kluge2024}. From the first all-sky survey of the western Galactic hemisphere, this catalog reports $12,247$ galaxy clusters with redshifts between 0.003 and 1.323 and extent likelihood \extlike$>3$. However, it is estimated to contain a contamination fraction of $>14\%$, primarily due to background fluctuations and misclassified AGN or other point-like sources \citep{Seppi2022, Clerc2024, Kluge2024, Balzer2025}.

\begin{figure*}
    \centering
    \begin{minipage}{0.45\textwidth}
        \centering
        \includegraphics[width=\textwidth, clip]{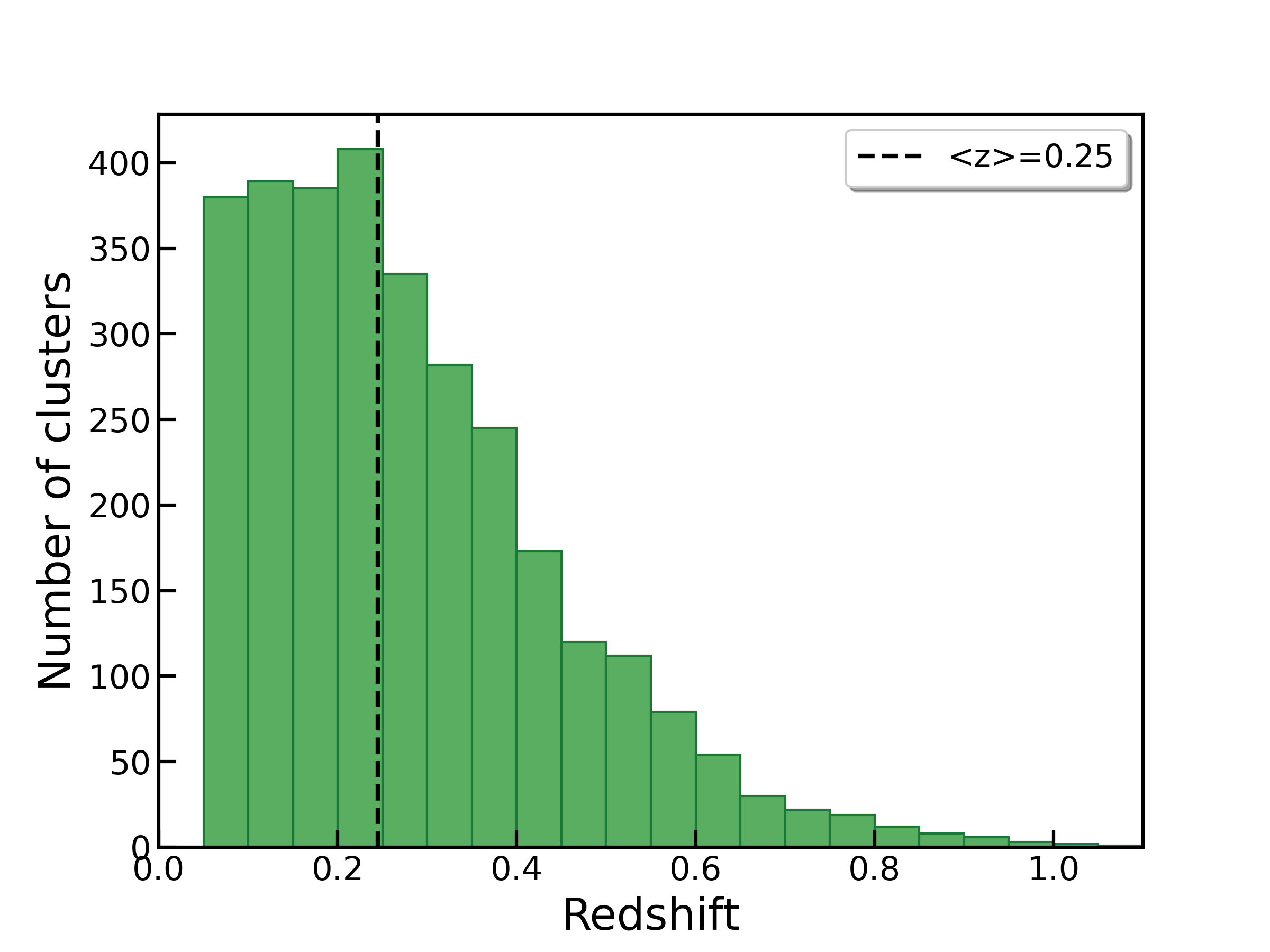}
    \end{minipage}
    \hspace{1cm}
    \begin{minipage}{0.45\textwidth}
        \centering
        \includegraphics[width=\textwidth, clip]{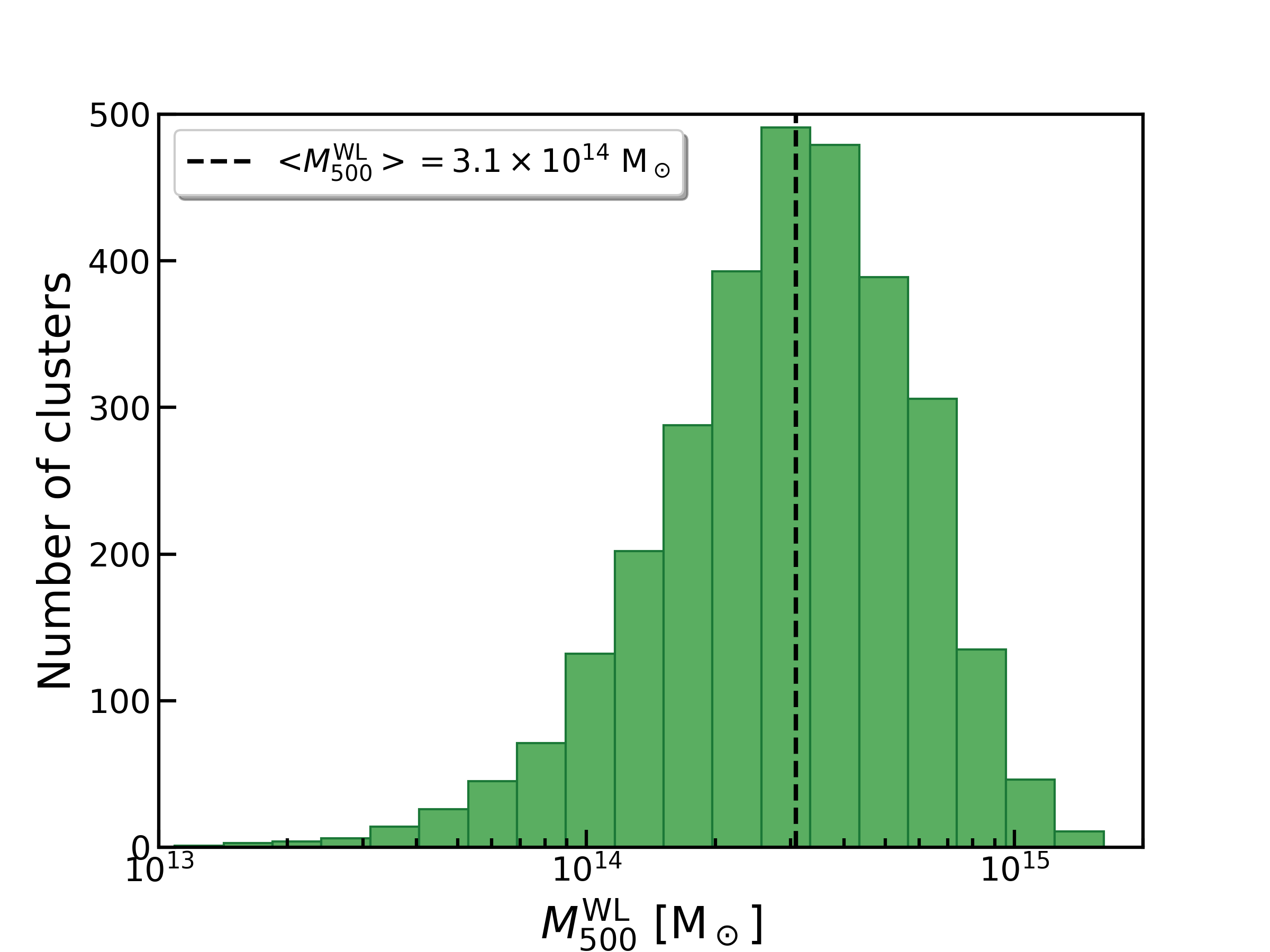}
    \end{minipage}
    \caption{Redshift ({\it left panel}) and mass distribution ({\it right panel}) of the $3061$ \erass\ clusters and groups used in this work. The median redshift and mass of the sample are shown by the vertical dashed black lines.}
    \label{fig:hist_z}
\end{figure*}

To utilize a clean sample in our analysis, we applied strict cuts on the \extlike$>12$, $z>0.05$, and $M_{500}\footnote{$M_{500}$ denotes the mass within $R_{500}$, defined as the radius enclosing a mean density equal to $500$ times the critical density of the Universe at the cluster’s redshift.}>1\times10^{13}$~M$_{\odot}$ (details of the mass determination are provided in Section~\ref{subsec:data_analysis}). Additionally, we applied a $\texttt{IN\_ZVLIM == TRUE}$ flag to avoid shallow optical/NIR data where the false-positive rate is known to increase steeply \citep{Kluge2024}. The final sample\footnote{Starting with $3414$ systems (\extlike$>12$), applying $z>0.05$ removes $146$, $M_{500}>1\times10^{13}$~M$_{\odot}$ removes $4$ more, and \texttt{IN\_ZVLIM = TRUE} excludes 203.} utilized in this work includes $3061$~clusters of galaxies and galaxy groups in the redshift range of $0.05<z<1.07$ with a median redshift of $0.25$. The purity level of the sample is estimated to be $\sim98\%$\footnote{The purity of the sample is obtained using Eq.~15 in \cite{Kluge2024}.}. This sample covers a mass range of $1.1\times10^{13}$~M$_{\odot}$ to $1.6\times10^{15}$~M$_{\odot}$, including $219$ galaxy groups, i.e., systems with $M_{500}$ below $1\times10^{14}$~M$_{\odot}$. This study includes the largest number of galaxy groups analyzed in the context of scaling relations. It comprehensively spans the mass range from low-mass galaxy groups to massive clusters. The redshift distribution of the sample is displayed in the left panel of Fig.~\ref{fig:hist_z}, and the $M_{500}$ cluster masses are shown on the right panel.

\subsection{Data reduction and analysis}
\label{subsec:data_analysis}

The X-ray properties of the clusters in our sample are extracted from \citet{Bulbul2024}, following an eROSITA standard data reduction and analysis procedure. The event files were processed using the eROSITA Science Analysis Software System \citep[eSASS;][]{Brunner2022}, with pipeline version 010 \citep{Merloni2024}. Images and exposure maps for each cluster were generated using the {\tt evtool} and {\tt expmap} commands. 

To obtain the physical properties of each cluster, \citet{Bulbul2024} performed imaging analysis using the MultiBand Projector 2D\footnote{ https://github.com/jeremysanders/mbproj2d} \citep[MBProj2D;][]{Sanders2018} tool. MBProj2D forward-fits cluster images across multiple bands using physical cluster models, including an electron density profile, a temperature profile, and, if the data quality allows, a metallicity profile. During this process, a Markov chain Monte Carlo (MCMC) analysis is used to generate posterior distributions of free parameters, such as the X-ray center, ICM temperature, and electron density model. 

When enough energy bands, particularly the hard band, are included in the fitting, MBProj2D can constrain the cluster temperature with comparable precision to spectral fitting. In the analysis of the eRASS1 clusters, \citet{Bulbul2024} used seven bands from $0.3$~keV to $7$~keV using parameter limits of $0.6<kT<25$~keV, and thus, were able to obtain temperature measurements for clusters when there were enough photons. Because most eRASS1 clusters have limited photon counts, they adopted an isothermal model for analysis. Therefore, only an average temperature is obtained for each cluster, rather than a temperature profile. Metallicity is fixed at $0.3~Z_{\odot}$ globally, with solar abundances adopted from \cite{Asplund2009}. The electron density profiles are described using the model from \citet{Vikhlinin2006a}, after discarding the second $\beta$ component, which is used to fit the cluster outskirts. The other quantities and their probability distribution, such as the soft-band X-ray luminosity (in the $0.5-2.0$~keV energy band) and gas mass, are derived from the chain of free parameters in the density profile and temperature. The full posterior distributions of the individual observables are explicitly incorporated into the scaling-relation fitting procedure (see Sect.~\ref{sec:like_fitting} for further details).

Due to the shallow nature of the eRASS1 survey, most eRASS1 clusters lack enough counts for precise temperature measurements. To assess temperature constraints, \cite{Bulbul2024} analyzed the MCMC chains, built logarithmic histograms of the temperature, and derived the marginal posterior distributions. Then, they computed the $98.7\%$ ($2.5\sigma$) highest density interval (HDI), reporting limits when the interval hits the parameter boundaries. In our sample of $3061$ clusters, $2$ have only upper limits, and $1663$ have lower limits on temperature. The remaining $1396$ clusters have constrained temperatures; the model-based posterior distribution is adopted as the best-fit temperature, and the $68.2\%$ HDI as the uncertainty range. For luminosity and gas mass, only $17$ clusters in our selected sample have upper-limit estimates, while the rest have constrained values. This does not pose a limitation for our scaling relation analysis, as the fitting procedure explicitly incorporates the full individual-observable posterior distributions. In doing so, the method treats measurement uncertainties in a statistically rigorous way and ensures that the resulting relations remain robust and reliable (see Sect.~\ref{sec:like_fitting}). Even if the analysis were restricted to clusters with well-constrained temperatures, for example, by increasing the $\mathcal{L}_{\mathrm{ext}}$  threshold, determining the selection function would not be straightforward. In the eRASS1 digital twin, only a limited number of detections exceed $\mathcal{L}_{\mathrm{ext}} > 12$, making it challenging to train the selection function (see Sect.~\ref{subsect:sel_func} and \citealt{Clerc2024}).

In this work, we used X-ray properties integrated over $R_{500}$, which are calculated from $M_{500}$ estimates. $M_{500}$ is inferred using the scaling relations between mass and X-ray count rate, with the calibration by weak lensing shear profiles \citep{Ghirardini2024, Grandis2024, Kleinebreil2024, Okabe2025}. This scaling relation is calibrated using the weak-lensing shear profiles from the Dark Energy Survey (DES), the Kilo Degree Survey (KiDS), and the HyperSuprime-Camera survey \citep{Okabe2025}.

\section{Scaling relation modeling}
\label{sec:scaling_modeling}

In this work, we followed the methods introduced by \citet{Bahar2022} to model the scaling relations of eROSITA-selected samples. To ensure a fair comparison between similarly selected samples, we employed the same scaling relations and likelihood-fitting method. We briefly discuss the method in this section. 

\subsection{Likelihood and fitting procedure}
\label{sec:like_fitting}

A simple scaling relation proposed by \citet{Kaiser1986} relies on the purely gravitational collapse model and is known as a self-similar relation. Assuming this model, one can derive the slope of the scaling relations; for example, the power-law index for the soft band X-ray luminosity ($L_{\mathrm{X}}$) and temperature ($T$) is $3/2$, while the slope is $1$ for the relation between $L_{\mathrm{X}}$ and gas mass ($M_{\mathrm{gas}}$).

We adopted the following general form of power-law relations between observables $Y$ and $X$, taking into account both redshift and mass evolution:
\begin{equation}\label{eq0}
\frac{Y}{Y_{\rm piv}} = A~\left (\frac{X}{X_{\rm piv}} \right )^B \left (\frac{E(z)}{E(z_{\rm piv})} \right )^C ,
\end{equation}
where $Y_{\rm piv}$ and $X_{\rm piv}$,  $z_{\rm piv}$ are the pivot values of the observables $Y$, $X$, and redshift of the sample. $A$ is the normalization, $B$ is the slope with respect to $X$, and $C$ is the slope of the redshift evolution of the power-law relation \citep{Bulbul2019, Bahar2022}. X-ray scaling relations are expected to evolve with redshift, even in a purely gravitational scenario, due to cosmological expansion and the resulting change in the Universe’s matter density. The redshift evolution is modeled using the evolution function, which is defined as $E(z) = H(z)/H_0$, where $H(z)$ is the Hubble parameter and $H_0$ is the Hubble constant. For all cases, we adopted a log-normal intrinsic scatter in the observables, $\sigma_{Y|X}$, defined in terms of the natural logarithm. For each scaling relation, we fit for four free parameters: $A,~B,~C,~\sigma_{Y|X}$. We also provided the best-fit scaling relations, assuming a self-similar evolution of the derived quantities in this work. 

\citet{Bahar2022} modeled the scaling relation between galaxy cluster observables by constructing a comprehensive likelihood function that incorporates selection effects, measurement uncertainties, and the underlying cosmological distribution of clusters. The full likelihood accounts for the detection probability of a cluster, the correlation in measurement uncertainties, and the intrinsic scatter in the scaling relation, all evaluated at a fixed cosmology. Given the high precision of spectroscopic and photometric redshift determinations ($\delta z/(1 + z) \leq 0.005$ in $0.05 < z < 0.9$) in the eRASS1 sample, we also treated the cluster redshifts as known and neglected their uncertainties in our analysis \citep{Kluge2024}. The instrumental effects, such as varying exposure time and telescope characteristics, are taken into account in the selection function (see Section~\ref{subsect:sel_func}).

As mentioned before, the observable–observable scaling relations are modeled as a log-normal distribution around a power-law form, and the cosmological distribution of the lower-level observable is derived by convolving the mass function with a mass–observable scaling relation calibrated via weak lensing. With these components, we marginalized over the true values of the observables to obtain the likelihood of the two measured  observables $X$ and $Y$ for a given cluster:
\begin{equation}\label{eq1}
\begin{split}
P(\hat{Y}, \hat{X}, I | \theta, z, S) = \iint_{Y,X} P(I | Y, z, S) \, P(\hat{Y}, \hat{X} | Y, X) \\\times \, P(Y | X, \theta, z) \, P(X | z) \, {\rm d}Y \, {\rm d}X \, ,
\end{split}
\end{equation}
where $P(I|Y, z, S)$ denotes the probability that a cluster with observable value $Y$, redshift $z$, and sky position $S$ is included in the sample, corresponding to the selection function derived from simulations (see Sect.~\ref{subsect:sel_func}). In contrast to \citet{Bahar2022}, and following \citet{Ghirardini2024} and \citet[][]{Sanders2025}, the sky position is explicitly included in the likelihood and selection function to account for the eRASS1 survey strategy, which produces position-dependent exposure times with deeper coverage near the ecliptic poles \citep{Predehl2021}. Additional sky-position-dependent effects, such as Galactic absorption \citep{HI4PI2016, Willingale2013}, detector response, and background levels \citep{Freyberg2020}, are also considered, as they affect the cluster detection efficiency and count-rate measurements. The term $P(\hat{Y}, \hat{X}|Y, X)$ describes the joint measurement uncertainties of the observables, including their correlation, and is constructed using the full MBProj2D posterior distributions of the individual observables (see Sect.~\ref{subsec:data_analysis}), thereby accounting for parameter degeneracies and the propagation of uncertainties between observables. The factor $P(Y|X, \theta, z)$ models the $Y$–$X$ scaling relation, while $P(X|z)$ represents the cosmological distribution of the observable $X$. The parameter vector $\theta$ contains the free parameters of the scaling relation, namely the normalization $A$, slope $B$, redshift evolution $C$, and intrinsic scatter $\sigma_{Y|X}$.

As in \citet{Bahar2022}, to mitigate the sensitivity to cosmological assumptions and the exact form of the mass-observable relation, we did not use the observed number of clusters as data. Instead, we treated the number of detections as a model-dependent quantity and constructed the cluster likelihood using Bayes' theorem. This approach allows us to focus on constraining the scaling relation parameters while avoiding terms that explicitly depend on cosmological volume or completeness corrections, thereby enhancing the robustness of our fits under a fixed cosmological model. This likelihood modeling is similar to the method presented in \cite{Giles2016} and \cite{Pacaud2016}. We derived the cosmological distribution of the observable $X$ as a function of redshift for a fixed cosmology ($P(X|z)$) and an assumed $X-M$ scaling relation. We transformed the \cite{Tinker2008} mass function into an “$X$-function.” This conversion is performed using the weak-lensing mass-calibrated scaling relations of \citet{Chiu2022}. The intrinsic scatter is explicitly incorporated into the modeling. First, we computed the expected value ($X_{\rm exp}$) from the \citet{Chiu2022} relation as a function of mass and redshift. We then modeled $X$ as a log-normal distribution centered on $X_{\rm exp}$, with a dispersion $\sigma_{\rm scat}$ corresponding to the scatter reported by \citet{Chiu2022}. Finally, the mass function is applied to derive the probability distribution $P(X|z)$ \citep[see ][for further details]{Bahar2022}.

Finally, we used the Bayes theorem to obtain the likelihood of the cluster X-ray observables given the probability of detection and model parameters,
\begin{equation}\label{eq2}
\mathcal{L}(\hat{Y_{i}}, \hat{X_{i}}|I, \theta, z_{i}, S_{i}) = \frac{P(\hat{Y_{i}}, \hat{X_{i}}, I | \theta, z_{i}, S_{i})}{\iint_{Y_{i}, X_{i}} P(I | Y, z, S) \, P(Y | X, \theta, z) \, P(X | z) \, {\rm d}Y_{i} \, {\rm d}X_{i}} .
\end{equation}
The final total likelihood is computed by multiplying the likelihoods of all individual clusters in the sample.

We fit the X-ray scaling relations using the {\tt emcee} MCMC sampler \citep{ForemanMackey2013} with the likelihood described above. We fit the eRASS1 cluster sample using flat priors: $\mathcal{U}(0.1,4)$ for the normalization $A$, $\mathcal{U}(-10,10)$ for the slope $B$, $\mathcal{U}(-10,10)$ for the redshift evolution $C$, and $\mathcal{U}(0.1,3.0)$ for the intrinsic scatter $\sigma_{Y|X}$. Pivot values correspond to the median of the observables (see Table~\ref{tab:pivval}). Each scaling relation is fit twice: once with $C$, the slope of the redshift evolution, as a free parameter, and once with $C$ fixed to the self-similar expectation. 

\begin{table}
\caption{Pivot values of the observables.}
\label{tab:pivval}
\centering
\begin{tabular}{c c}
\hline\hline
X-ray observable & Median value \\
\hline
    $z$           & $0.25$ \\
    $L_{\rm X}$   & $6.53 \times 10^{43}$~erg~s$^{-1}$ \\
    $T$           & $2.27 $~keV \\
    $M_{\rm gas}$ & $ 2.28\times10^{13}$~M$_{\odot}$ \\
    $Y_{\rm X}$   & $ 4.94\times10^{13}$~M$_{\odot}$~keV \\
\hline
\end{tabular}
\tablefoot{The values represent the sample median.}
\end{table}

\subsection{Selection function}
\label{subsect:sel_func}

Proper scaling relation fitting depends heavily on accurate knowledge of selection effects, such as survey incompleteness and contamination. The selection function model $P(I|Y, z, S)$, adopted here, closely follows the approach of \cite{Ghirardini2024} and is based on multiple mock realizations of the eROSITA all-sky survey \citep{Seppi2022, Clerc2024}. These simulations are designed to replicate the key instrumental characteristics of eROSITA, including exposure inhomogeneities, point-spread function variations, effective area changes, and the grasp of all seven telescopes. Foreground and background sources, such as stars, AGN, and galaxy clusters, are embedded within a full-sky light-cone N-body simulation \citep{Comparat2020}. AGN properties are assigned using abundance-matching techniques, while the emissivity profiles of galaxy clusters and groups are drawn from an observational template library, with associations based on halo mass, redshift, and dynamical state.

Source detection in the mock eRASS1 data follows the procedure detailed in \cite{Seppi2022}. Detected sources are matched to simulated ones in the sky by considering spatial proximity, extent, and brightness, favoring associations with the brightest candidate in cases of ambiguity. Successful matches are flagged as true detections. This process produces a labeled dataset from which detection probabilities can be estimated across a broad range of cluster properties.

The modeling of detection probabilities is performed using Gaussian Process (GP) classifiers, trained on two-thirds of the matched simulated sample \citep{Clerc2024}. These nonparametric models interpolate the detection probability across the multidimensional space defined by cluster properties, such as soft-band luminosity, redshift, and local exposure, optionally including central emission measure. The remaining parameters are marginalized over, assuming the simulations capture their statistical distributions. Compared to spline interpolation, GP classifiers offer a more rigorous handling of uncertainty, particularly in sparsely sampled regions of parameter space. Model calibration is verified by checking that the predicted detection probabilities agree with the actual detection rates on the test set.

The selection function used in this work follows the cluster selection described in Sect.~\ref{subsec:sample_selection}. It includes extended sources with $\mathcal{L}_{\mathrm{ext}} > 12$, a simulated redshift range of $0 \leq z \leq 1.5$, and a lower mass limit of $M_{500}=1\times10^{13}$~M$_{\odot}$. In addition, sources are required to satisfy {\tt IN\_XGOOD==True}, a spatial mask from the optical identification that is connected to the {\tt IN\_ZVLIM} flag (see Sect.~\ref{subsec:sample_selection} and \citealt{Kluge2024}).

\begin{figure}
    \centering
    \includegraphics[width=0.9\linewidth]{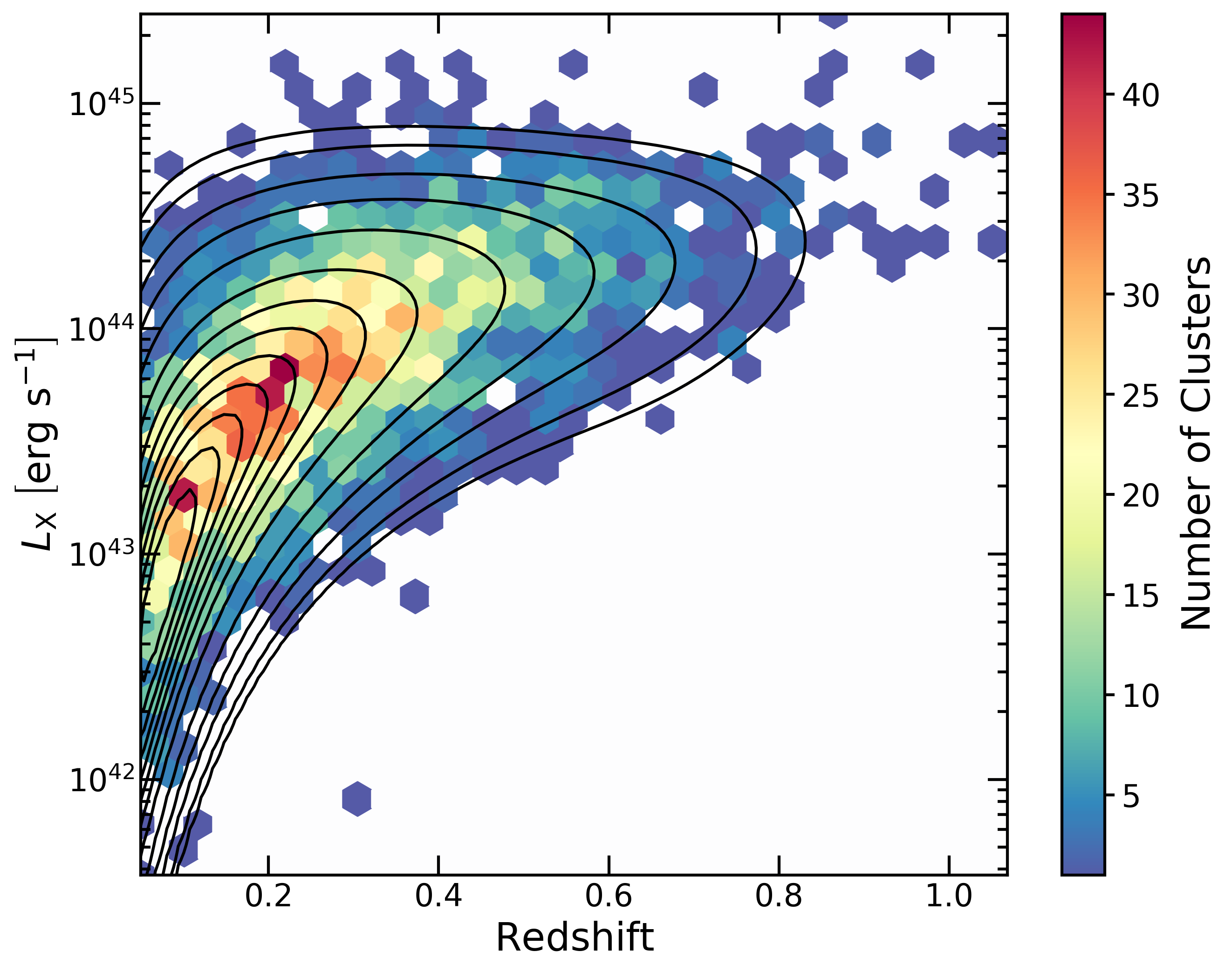}
    \caption{Hexbin plot of the soft band X-ray luminosity ($0.5-2.0$~keV energy band) and redshift distribution of the selected $3061$ eRASS1 clusters. Solid black curves show the PDF of the hypothetical $L_{\rm X}-z$ distribution given by the selection function (see Section~\ref{subsect:sel_func} for details).}
    \label{fig:Lx_z_distribution}
\end{figure}

As in \cite{Bahar2022}, to evaluate the performance and realism of the selection function, we compared the modeled luminosity distribution of detected clusters with observations. Specifically, we computed $P(I, L_{\mathrm{X}}, z, S) = P(I|L_{\mathrm{X}}, z, S) P(L_{\mathrm{X}}|z) P(z)$, where $P(L_{\mathrm{X}}|z)$ is obtained using Eq.~6 in \cite{Bahar2022} together with the best-fit $L_{\mathrm{X}}\text{--}M$ relation from \cite{Chiu2022} and the \cite{Tinker2008} mass function; and $P(z)$ is proportional to the comoving volume element d$V_c(z)$, assuming a constant comoving cluster density in the redshift range $0<z<1.1$. The comparison, shown in Fig.~\ref{fig:Lx_z_distribution}, demonstrates strong agreement between the observed (hexbin plot) and modeled (contours) soft band X-ray luminosity--redshift distributions, supporting the reliability and representativeness of the selection function within the eRASS1 survey context.

\subsection{Comparison of scaling relations with literature}
\label{subsect:comparison}

A careful comparison between scaling relations in the literature and our results requires standardizing several methodological choices, including the adopted functional forms, energy bands, cosmology, etc. To ensure consistency, we applied well-defined corrections before comparing.

We adopted self-similar redshift evolution as a common reference frame and converted all observables accordingly; for example, luminosities are measured in the $0.5-2.0$~keV energy band. For literature scaling relations reported in the $0.1-2.4$~keV energy band, we computed a conversion factor using an unabsorbed {\tt apec} model in {\tt XSPEC} \citep{Arnaud1996}, assuming a cluster temperature of $2.27$~keV, metallicity of $0.3$, and redshift of $0.25$, corresponding to the median properties of our sample. This yields a conversion factor of $1.63$. Variations in temperature and redshift modify this factor by only a few percent, and we therefore applied this factor uniformly to studies using the $0.1-2.4$~keV energy band. Finally, all relations are converted to a dimensionless Hubble constant of $0.6774$, matching the cosmology adopted in this work. These adjustments affect only the normalizations, while the published slopes and redshift evolution parameters remain unchanged.

Overall, comparisons with literature results should be treated with some caution. Earlier studies often did not account for selection effects, and the mass and redshift ranges probed can differ between works. Both factors can lead to variations in the inferred normalization and slope of the scaling relations, and likely contribute to the observed discrepancies.


%
\section{Results}
\label{sec:results}
In this section, we study the scaling relations between X-ray observables of securely detected eRASS1 clusters of galaxies. As mentioned in Section~\ref{sec:intro}, X-ray observable scaling relations help understand the physics of the intracluster medium in galaxy groups and clusters. This work shows the $L_{\mathrm{X}}-T$, $L_{\mathrm{X}}-M_{\mathrm{gas}}$, $L_{\mathrm{X}}-Y_{\mathrm{X}}$, and, $M_{\mathrm{gas}}-T$ scaling relations. A comparison with previous scaling relation results is also shown. 

In contrast with \cite{Bahar2022}, we did not include scaling relations that involve the bolometric luminosity. The reason is that, using eROSITA data, we measured the soft-band luminosity directly, as determined in the standard eROSITA $0.2-2.3$~keV energy band. The conversion from soft band to bolometric luminosity depends highly on the temperature and its uncertainty. Moreover, the selection function ultimately depends on the distribution of photon events in the soft band. Since the selection function is calibrated using simulations, all steps linking the soft-band event distribution to the bolometric luminosity are inherently model-dependent. Consequently, the selection function is sensitive to the assumptions underlying the digital twin, further adding to the analysis's uncertainty.

Table~\ref{tab:best_fit_parameters} presents the best-fit results for all scaling relations from both analyses, considering cases with the slope of the redshift evolution, $C$, left free, as well as fixed to the self-similar value. Appendix~\ref{app:parameter_constraints} provides the marginalized and joint posterior distributions of the model parameters for all scaling relation fittings.

\begin{table*}
\caption{Best-fit parameters of the scaling relations.}
\label{tab:best_fit_parameters}
\centering
\renewcommand{\arraystretch}{1.4}
\begin{tabular}{c c | >{\centering\arraybackslash}m{14mm} >{\centering\arraybackslash}m{14mm} >{\centering\arraybackslash}m{14mm} >{\centering\arraybackslash}m{14mm} | >{\centering\arraybackslash}m{14mm} >{\centering\arraybackslash}m{14mm} >{\centering\arraybackslash}m{14mm} >{\centering\arraybackslash}m{14mm}}
\hline\hline
\multicolumn{2}{c}{Scaling relation} & \multicolumn{4}{c}{Free redshift evolution} & \multicolumn{4}{c}{Self-similar redshift evolution} \\
$Y$ & $X$ & $A$ & $B$ & $C$ & $\sigma_{Y|X}$ & $A$ & $B$ & $C_{\mathrm{SS}}$ & $\sigma_{Y|X}$ \\
\hline
    $L_{\rm X}$   & $T$           & $0.41^{+0.02}_{-0.02}$ & $2.51^{+0.04}_{-0.04}$ & $1.88^{+0.34}_{-0.37}$ & $ 0.57^{+0.03}_{-0.03}$ & $ 0.39^{+0.02}_{-0.02}$ & $ 2.54^{+0.04}_{-0.03}$ & $1$ & $ 0.60^{+0.03}_{-0.03}$ \\
    $L_{\rm X}$   & $M_{\rm gas}$ & $ 0.988^{+0.006}_{-0.006}$ & $ 1.234^{+0.006}_{-0.006}$ & $ 2.16^{+0.07}_{-0.07}$ & $ 0.188^{+0.005}_{-0.005}$ & $ 0.988^{+0.005}_{-0.005}$ & $ 1.241^{+0.006}_{-0.006}$ & $2$ & $ 0.190^{+0.005}_{-0.005}$ \\
    $L_{\rm X}$   & $Y_{\rm X}$   & $ 0.86^{+0.01}_{-0.01}$ & $ 0.855^{+0.009}_{-0.009}$ & $ 2.09^{+0.16}_{-0.17}$ & $ 0.25^{+0.01}_{-0.01}$ & $ 0.85^{+0.01}_{-0.01}$ & $ 0.866^{+0.008}_{-0.008}$ & $8/5$ & $ 0.27^{+0.01}_{-0.01}$ \\
\hline
    $M_{\rm gas}$ & $T$           & $ 0.44^{+0.02}_{-0.02}$ & $ 2.24^{+0.03}_{-0.03}$ & $ -1.00^{+0.29}_{-0.30}$ & $ 0.49^{+0.02}_{-0.02}$ & $ 0.44^{+0.02}_{-0.02}$ & $ 2.22^{+0.03}_{-0.03}$ & $-1$ & $ 0.48^{+0.02}_{-0.02}$ \\
\hline
\end{tabular}
\tablefoot{
The fitted relation is of the form $\frac{Y}{Y_{\rm piv}} = A~\left (\frac{X}{X_{\rm piv}} \right )^B \left (\frac{E(z)}{E(z_{\rm piv})} \right )^C$, with a log-normal scatter of $\sigma_{Y|X}$ (defined in terms of the natural logarithm). The fitting is performed twice: with a free redshift evolution (exponent $C$), and with a redshift evolution fixed to the self-similar expectation. The displayed errors are $1\sigma$ uncertainties obtained from the MCMC chains.
}
\end{table*}

\begin{figure}
    \centering
    \includegraphics[width=\linewidth]{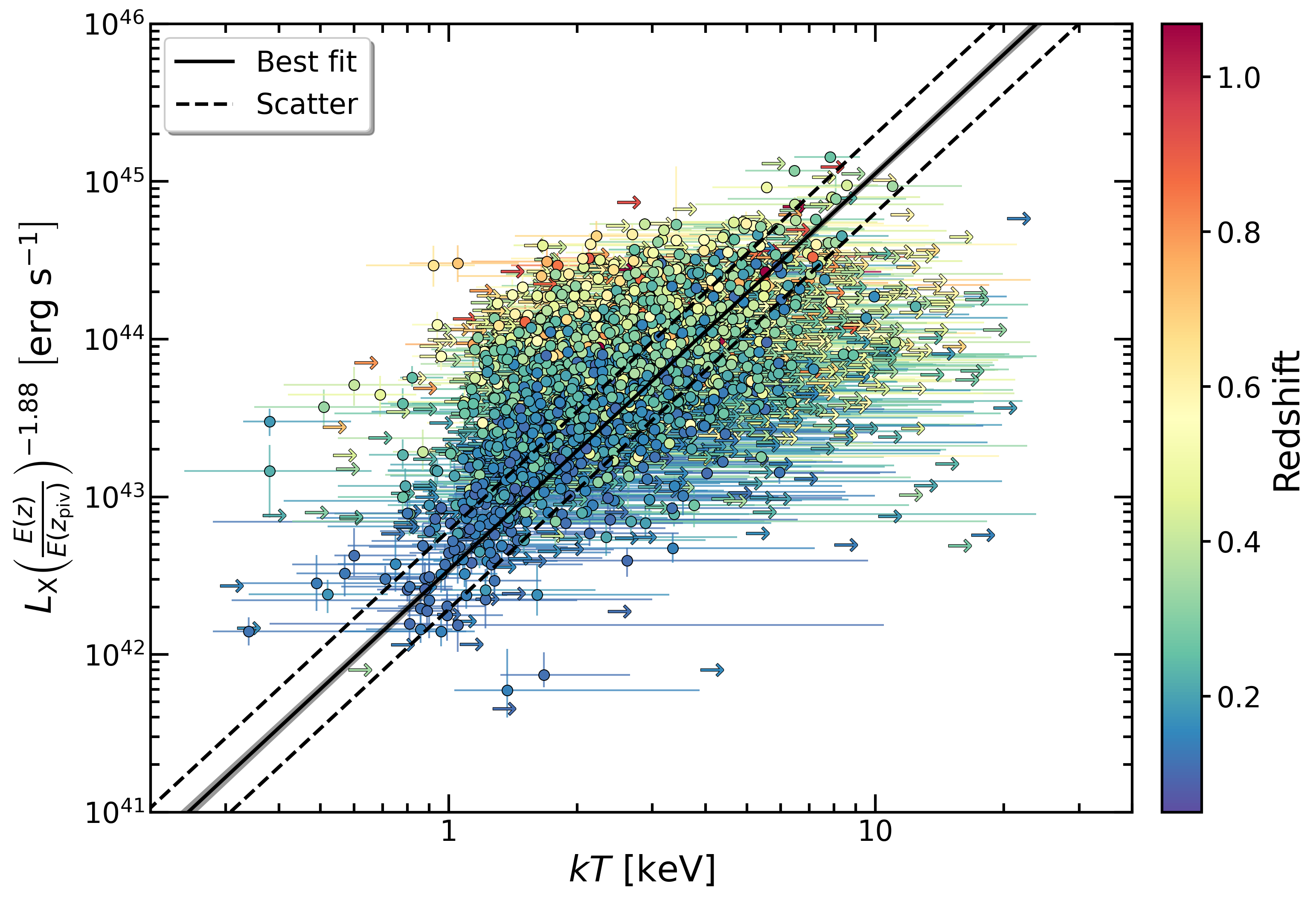}
    \caption{Soft band ($0.5-2.0$~keV energy band) X-ray luminosity--temperature, $L_{\rm X}-T$, scaling relation of the selected galaxy group and cluster eRASS1 sample. Data points are color-coded by redshift. Arrows indicate temperature lower limits for objects in the sample with poorly constrained temperatures. The black line shows the best-fit scaling relation model. The shaded black area indicates the $1\sigma$ uncertainty of the mean of the log-normal model, and the dashed black lines indicate the best-fit standard deviation ($\sigma_{L_{\rm X}|T}$) around the mean.}
    \label{fig:Lx_T_best_fit_model}
\end{figure}

\begin{figure}
    \centering
    \includegraphics[width=0.8\linewidth]{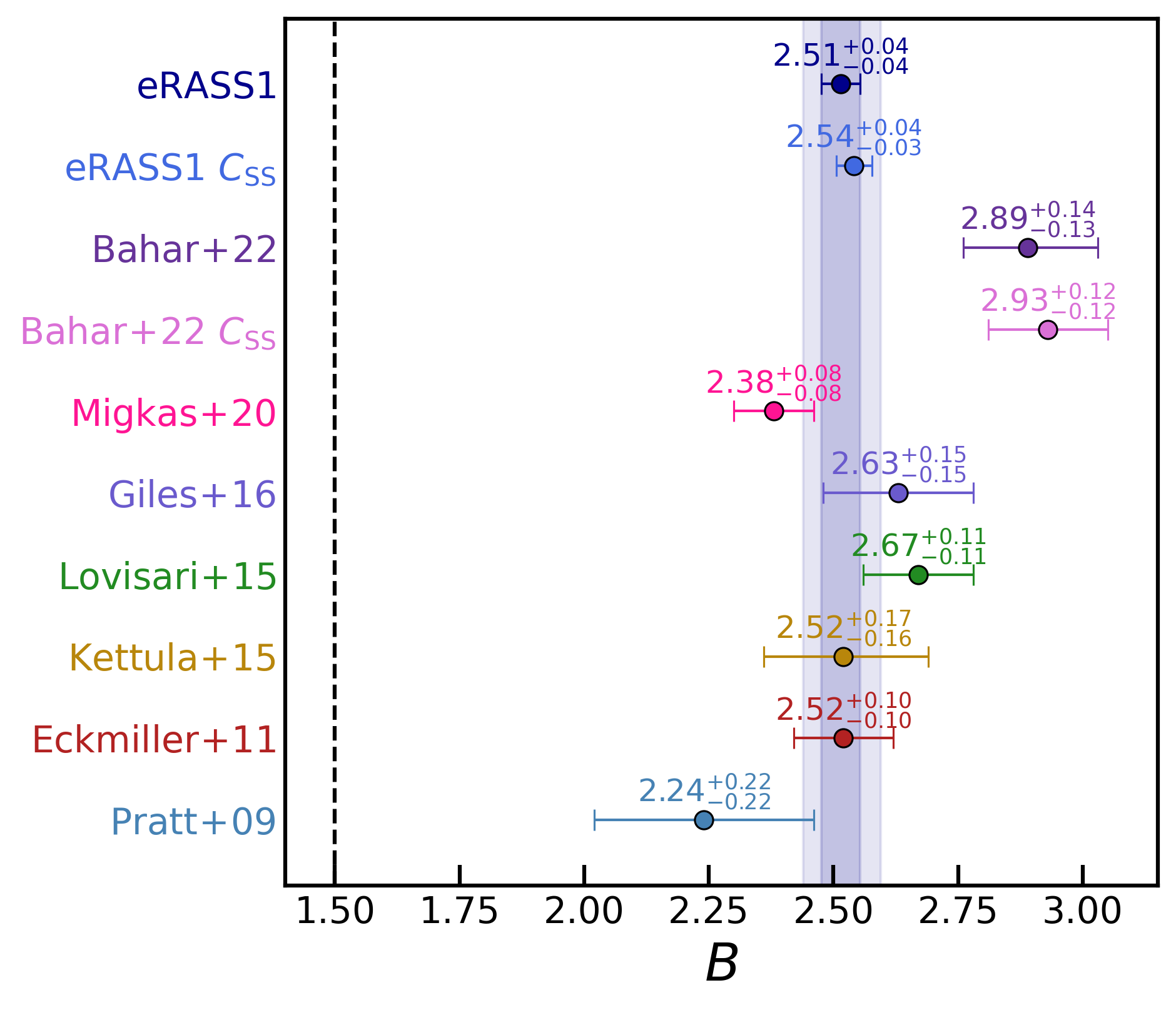}
    \caption{Comparison of the slope $B$ for the $L_{\rm X}-T$ scaling relation with literature results from \citet[eFEDS cluster sample]{Bahar2022}, \cite{Migkas2020}, \citet[XXL cluster sample]{Giles2016}, \cite{Lovisari2015}, \cite{Kettula2015}, \cite{Eckmiller2011}, and \cite{Pratt2009}. All the shown errors are at the $1\sigma$ level. The vertical shaded areas represent our best-fit measurements at the $1-2\sigma$ confidence level. The vertical dashed black line displays the self-similar expected value.}
    \label{fig:Lx_T_B_comparison}
\end{figure}

\subsection{Soft band X-ray luminosity--temperature scaling relation}
\label{subsec:lx_t_rel}

The soft band X-ray luminosity--temperature, $L_{\mathrm{X}}-T$, relation is one of the most extensively studied among X-ray scaling relations, due to the relatively straightforward and almost independent measurement of both quantities using X-ray data. Both quantities probe different yet intrinsically connected aspects of the ICM: luminosity scales with the square of the electron density, making it highly sensitive to the spatial distribution of the hot gas, while temperature reflects the average kinetic energy of the ICM.

Previous studies revealed that the observed $L_{\mathrm{X}}-T$ relation does not follow the expected self-similar scaling ($L_{\mathrm{X}}\propto T^{3/2}$) derived from purely gravitational heating processes \citep[; e.g.,][]{Markevitch1998, Ettori2004, Pratt2009, Eckmiller2011, Maughan2012, Hilton2012, Kettula2015, Lovisari2015, Giles2016, Bahar2022}. Although the precise slope varies across studies due to factors such as modeling approaches, sample selection, and calibration, there is a general consensus that the relation has a slope between $2.3$ and $3.0$. Numerous observational studies have consistently confirmed this deviation and attributed it to additional, nongravitational physical processes \citep[for reviews see][]{Giodini2013, Lovisari2022}. Therefore, a deeper understanding of the $L_{\mathrm{X}}-T$ relation can offer valuable insights into the thermal history of galaxy clusters.

\begin{figure*}
    \centering
    \includegraphics[width=0.49\linewidth]{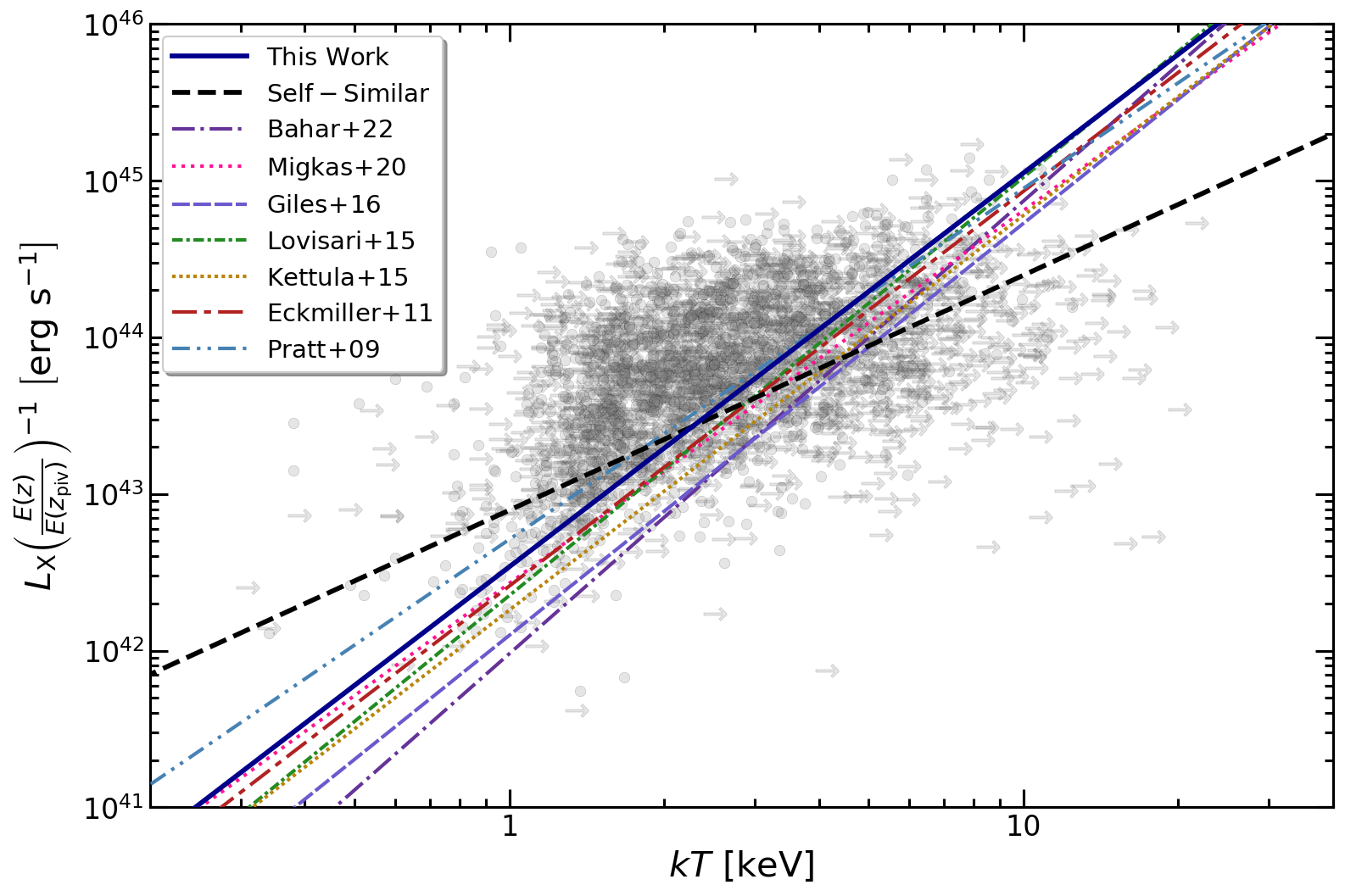}
    \includegraphics[width=0.49\linewidth]{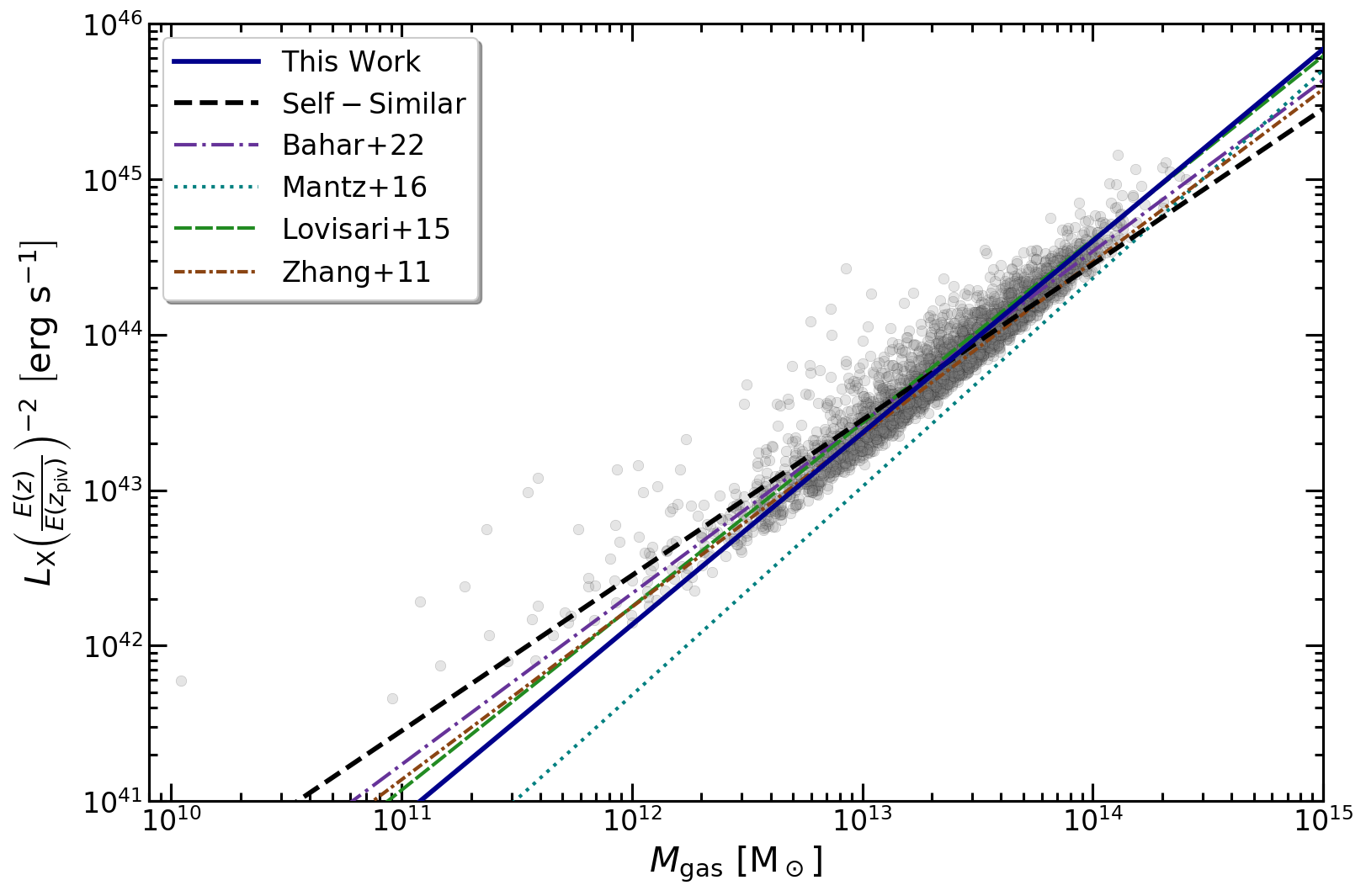}
    \includegraphics[width=0.49\linewidth]{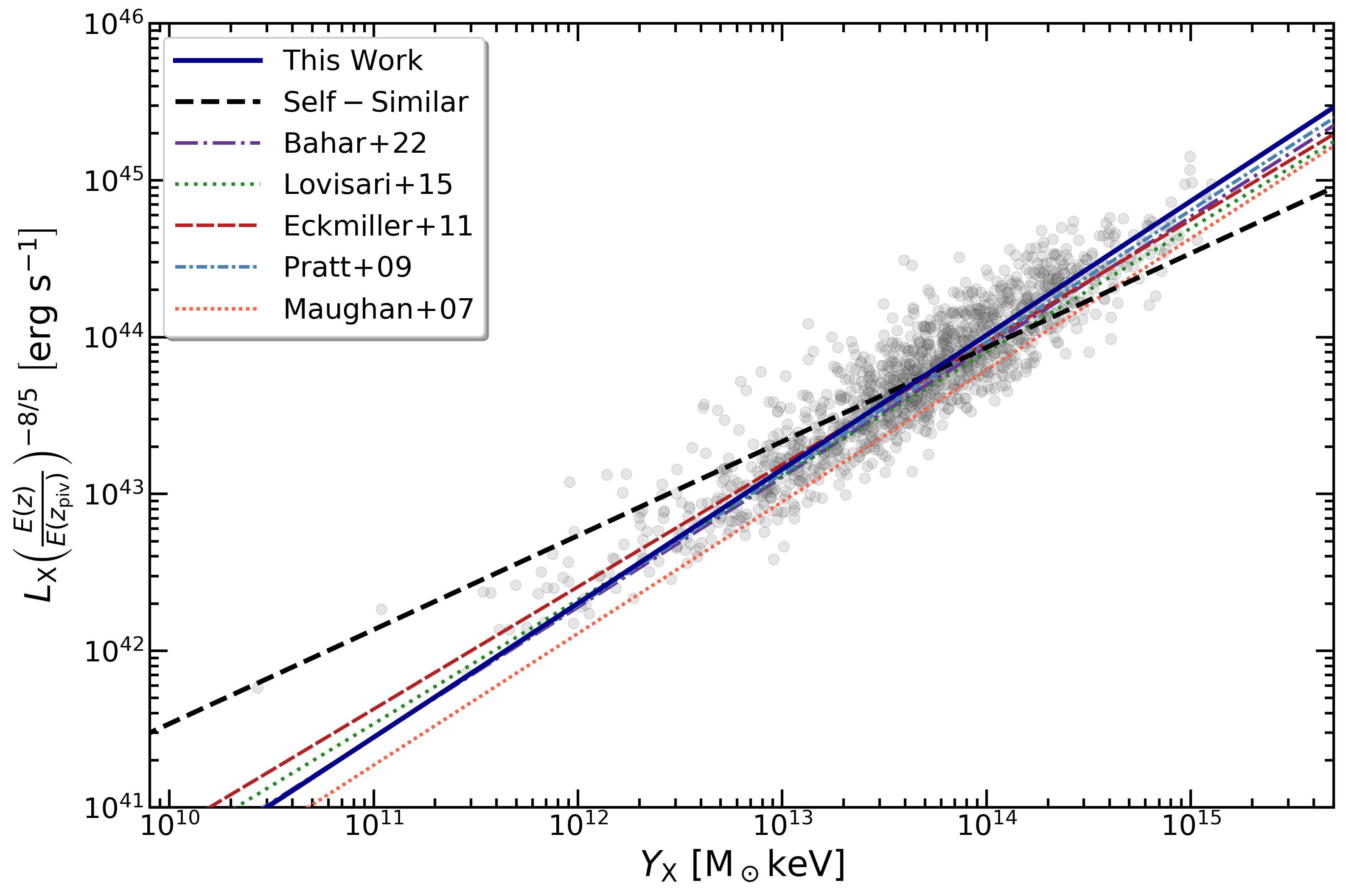}
    \includegraphics[width=0.49\linewidth]{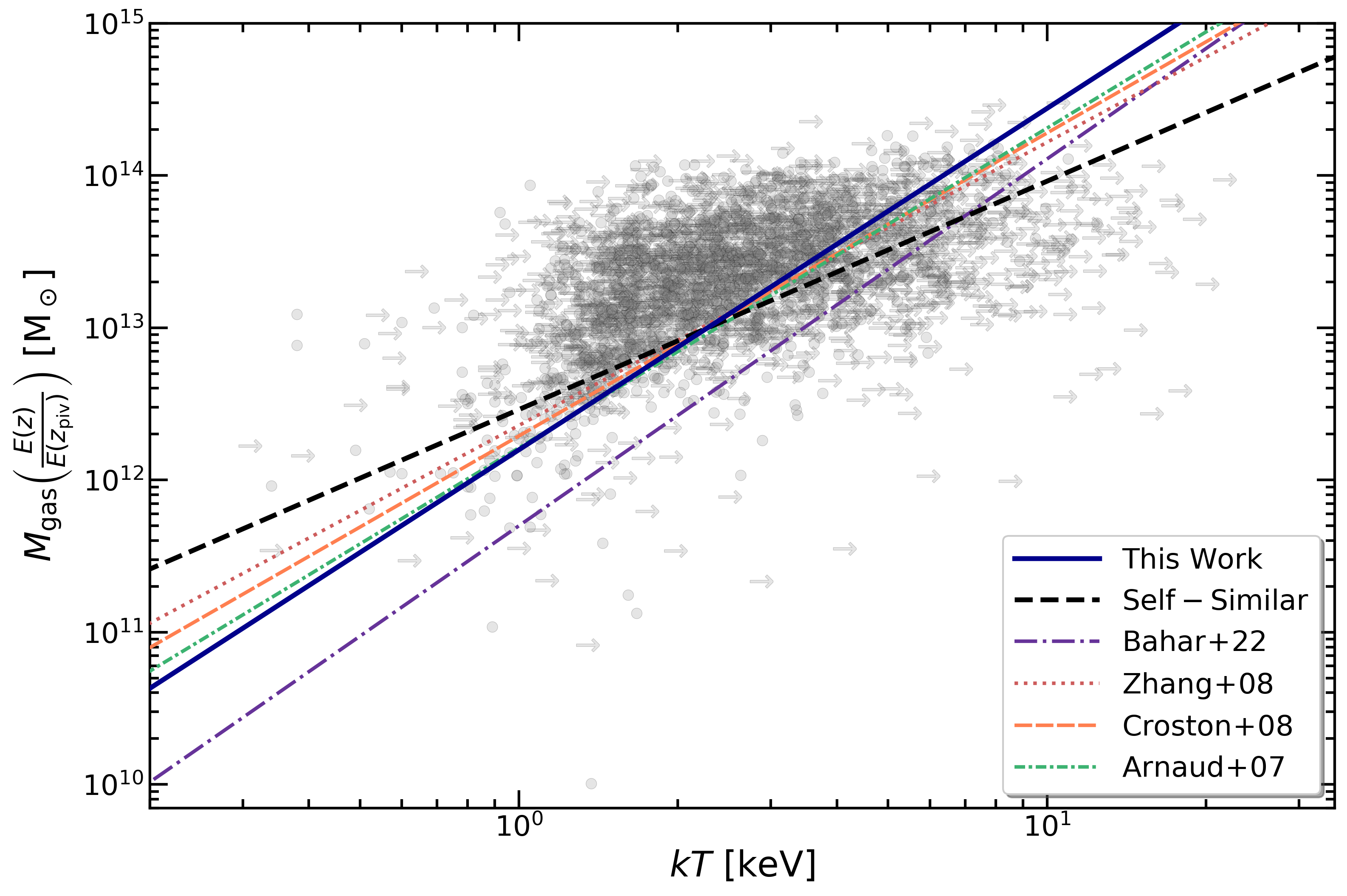}
    \caption{Comparison of our best-fit scaling relations, $L_{\rm X}-T$ ({\it top left panel}), $L_{\rm X}-M_{\rm gas}$ ({\it top right panel}), $L_{\rm X}-Y_{\rm X}$ ({\it bottom left panel}), and $M_{\rm gas}-T$ ({\it bottom right panel}), with predictions from the self-similar model and results from previous studies in the literature \citep{Arnaud2007, Maughan2007, Croston2008, Zhang2008, Pratt2009, Eckmiller2011, Zhang2011, Kettula2015, Lovisari2015, Giles2016, Mantz2016, Migkas2020, Bahar2022}. Gray symbols represent our eRASS1 galaxy cluster sample. Error bars have been omitted for visual clarity. To ensure consistency, all literature results have been converted to a common cosmology and energy band (see Sect.~\ref{subsect:comparison} for details).}
    \label{fig:scaling_rel_comparison}
\end{figure*}

Our best-fit parameters for the $L_{\rm X}-T$ scaling relation are: a slope of $B=2.51^{+0.04}_{-0.04}$, a redshift evolution parameter of $C=1.88^{+0.34}_{-0.37}$, a normalization of $A=0.41^{+0.02}_{-0.02}$, and an intrinsic scatter of $\sigma_{L_{\rm X}|T}=0.57^{+0.03}_{-0.03}$. Fig.~\ref{fig:Lx_T_best_fit_model} shows the corresponding best-fit model. Our best-fit slope, $B$, significantly deviates from the expected self-similar value ($\sim 25\sigma$). However, it is consistent with most previous studies within $1\sigma$ (see Fig.~\ref{fig:Lx_T_B_comparison}), except for the result from \citet{Bahar2022}, $B=2.89^{+0.14}_{-0.13}$, which shows a discrepancy at the $\sim2.6\sigma$ level. The group and cluster sample used in this work is significantly larger than those in previous studies, resulting in much smaller uncertainties on our best-fit parameters and placing tight constraints on them. Our results show that the redshift evolution parameter is consistent with the self-similar expectation for temporal evolution ($C_{\rm SS}=1$), showing an agreement at the $\sim2.4\sigma$ level. \cite{Giles2016}, analyzing the 100 brightest clusters in the XXL Survey, reported $C=1.64^{+0.77}_{-0.77}$, which is consistent with our result.

Fixing the evolution parameter to its self-similar value yields a slope of $B=2.54^{+0.04}_{-0.03}$, which is fully consistent with the case where evolution is left free. However, this slope is shallower than the value reported by \cite{Bahar2022}, $B=2.93^{+0.12}_{-0.12}$. The intrinsic scatter best-fit obtained in this work agrees well within $2\sigma$ with the ones published previously by different authors, except for the ones from \cite{Bahar2022}. Fig.~\ref{fig:Lx_T_sigma_comparison} shows a comparison of the intrinsic scatter $\sigma_{L_{\rm X}|T}$ with literature results. Our $\sigma_{L_{\rm X}|T}$ results agree within $1\sigma$, with previous studies \citep[e.g.,][]{Pratt2009, Eckmiller2011, Lovisari2015, Giles2016}, however, it is shallower than the value obtained by \cite{Bahar2022}.

The upper-left panel of Fig.~\ref{fig:scaling_rel_comparison} compares our best-fit relation with results from the literature, following the method described in Sect.~\ref{subsect:comparison}. The relation of \cite{Pratt2009}, derived from massive, low-redshift clusters ($M_{500}>10^{14}$~M$_{\odot}$), deviates from ours at low temperatures. In contrast, studies based on lower-mass systems ($M_{500}<10^{14}$~M$_{\odot}$; \citealt{Eckmiller2011, Lovisari2015, Kettula2015}) yield a lower normalization. Our higher normalization appears driven by clusters around $\sim2$~keV with relatively high luminosities ($\gtrsim10^{43.5}$~erg~s$^{-1}$), a population that differs from the XXL sample analyzed by \cite{Giles2016}. The discrepancy with \citet{Bahar2022} may be attributed to the larger fraction of galaxy groups in their sample. Several studies have found that the slope $B$ of the $L_{\rm X}-T$ scaling relation of galaxy groups (systems with $kT<3.5$~keV) is steeper than the one for galaxy clusters, a difference attributed to their shallower potential wells and the stronger influence of nongravitational processes \citep[e.g.,][]{Maughan2012}. In \cite{Bahar2022}, $26\%$ of their sample are groups with masses below $<10^{14}$~M$_\odot$, compared to only $\sim7\%$ in our analysis. This difference arises from the deeper eFEDS data used in their study, which enable more effective detection of lower-mass systems than the eRASS1 observations used here. Therefore, they obtain a steeper slope, potentially reflecting underlying sample-selection effects. Moreover, the sample of \cite{Bahar2022} includes low-luminosity systems ($\sim10^{41}$~erg~s$^{-1}$) that are not present in our data, while our sample extends to higher luminosities. This difference in luminosity coverage may contribute to their lower normalization relative to our best-fit relation. In Section~\ref{subsec:galgroups_discussion}, we discuss further the impact of galaxy groups on the scaling relation fitting.

\subsection{Soft band X-ray luminosity--gas mass scaling relation}
\label{subsec:lx_gmass_relation}

X-ray luminosity and gas mass are intrinsically correlated observables, primarily because they depend on the intracluster and intragroup electron density. As a result, a strong statistical correlation is expected between them. A precise characterization of this correlation must account for selection biases, the underlying halo mass function, and use a statistically significant sample.

Most previous studies of the $L_{\mathrm{X}}-M_{\rm gas}$ relation show a slight deviation from the self-similar expectations with respect to $M_{\rm gas}$ ($L_{\mathrm{X}}\propto M_{\rm gas}$). The results vary across studies, with differences attributed to variations in modeling approaches, sample selection, and calibration \citep[e.g.,][]{Zhang2011, Pratt2009, Lovisari2015, Mantz2016, Bahar2022}.

\begin{figure}
    \centering
    \includegraphics[width=\linewidth]{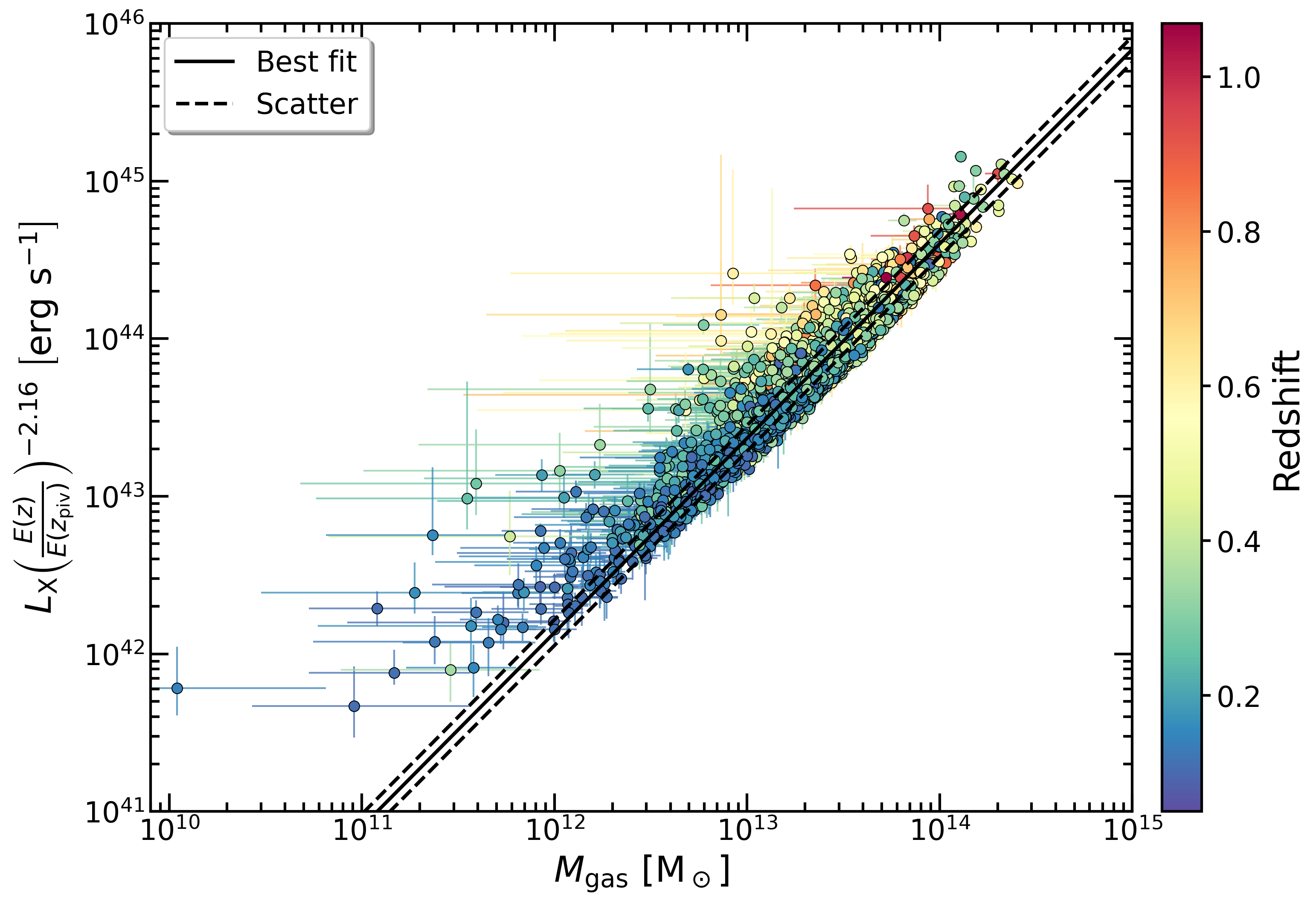}
    \caption{Same as Fig.~\ref{fig:Lx_T_best_fit_model} but for the soft band ($0.5-2.0$~keV energy band) X-ray luminosity--gas mass, $L_{\rm X}-M_{\rm gas}$, scaling relation.}
    \label{fig:Lx_Mgas_best_fit_model}
\end{figure}

Our best-fit parameters for the $L_{\rm X}-M_{\rm gas}$ scaling relation are: a slope of $B=1.234^{+0.006}_{-0.006}$, a redshift evolution parameter of $C=2.16^{+0.07}_{-0.07}$, a normalization of $A=0.988^{+0.006}_{-0.006}$, and an intrinsic scatter of $\sigma_{L_{\rm X}|M_{\rm gas}}=0.188^{+0.005}_{-0.005}$. This best-fit model is presented in Fig.~\ref{fig:Lx_Mgas_best_fit_model}. Our best-fit slope is steeper and statistically inconsistent with the self-similar expectation ($\sim 23\sigma$), while the redshift evolution parameter slope is consistent with the self-similar model ($C_{\rm SS}=2$) at $\sim 2\sigma$. The significantly larger sample considered in this work allows for tighter constraints on the $L_{\rm X}-M_{\rm gas}$ relation. By fixing the redshift evolution exponent to the self-similar value, the slope does not significantly change, $B=1.241^{+0.006}_{-0.006}$.

\begin{figure}
    \centering
    \includegraphics[width=0.8\linewidth]{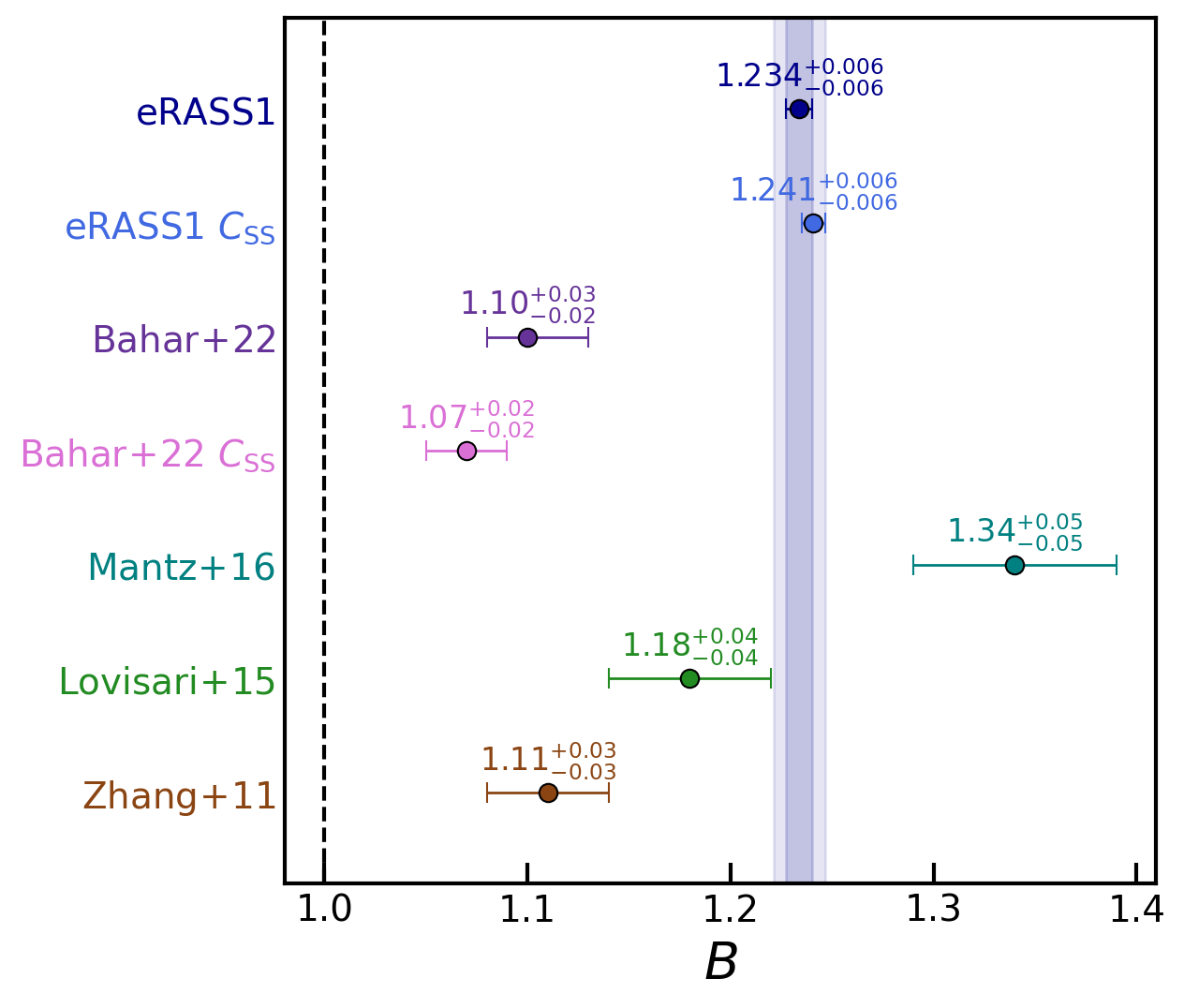}
    \caption{Same af Fig.~\ref{fig:Lx_T_B_comparison} but showing the slope $B$ comparison of the $L_{\rm X}-M_{\rm gas}$ scaling relation, including results from \citet[eFEDS cluster sample]{Bahar2022}, \cite{Mantz2016}, \cite{Lovisari2015}, and \cite{Zhang2011}.}
    \label{fig:Lx_Mgas_B_comparison}
\end{figure}

Fig.~\ref{fig:Lx_Mgas_B_comparison} compares our best-fit slope results with previous studies, considering both the self-similar and nonself-similar cases. Our findings are consistent within $2\sigma$ with studies that include high-mass systems; for example, \citet{Mantz2016}, which include massive, luminous clusters, and report slopes steeper than the self-similar expectation. In contrast, studies that focus on or include a significant fraction of galaxy groups tend to find shallower slopes \citep[e.g.,][]{Zhang2011, Lovisari2015, Bahar2022}. This trend is also evident in the group-only analysis by \citet{Lovisari2015}, who report a slope of $B=1.18^{+0.04}_{-0.04}$. The upper-left panel of Fig.~\ref{fig:scaling_rel_comparison} also compares our best-fit results with those reported in previous studies. Overall, the normalization is consistent across the different works, except for \cite{Mantz2016}. This discrepancy is likely due to the absence of low-mass systems in their sample.

As noted by \cite{Bahar2022}, only a few studies have reported the intrinsic scatter of the $L_{\rm X}-M_{\rm gas}$ relation. Fig.~\ref{fig:Lx_Mgas_sigma_comparison} presents a comparison between our measured $\sigma_{L_{\rm X}|M_{\rm gas}}=0.188^{+0.005}_{-0.005}$ and values from the literature. We found significant tension with previous results, at the $\sim 2.3\sigma$ level compared to \cite[][$\sigma_{L_{\rm X}|M_{\rm gas}}=0.14^{+0.02}_{-0.02}$]{Zhang2011}, and at the $\sim 5.4\sigma$ level relative to \cite[][$\sigma_{L_{\rm X}|M_{\rm gas}}=0.30^{+0.02}_{-0.02}$]{Bahar2022}. This may be attributed to differences in the sample sizes of galaxy groups used across the works, where feedback mechanisms can play a significant role.

\subsection{Soft band X-ray luminosity--low-scatter mass proxy scaling relation}

\cite{Kravtsov2006} introduced the X-ray analog of the integrated Compton-$Y$ parameter, defined as $Y_{\rm X}=M_{\rm gas}\times T$. This quantity is linked to the ICM's total thermal energy and serves as a low-scatter mass proxy. Numerical simulations have demonstrated that $Y_{\rm X}$ is insensitive to the specific assumptions used in modeling feedback processes \cite[e.g.,][]{Stanek2010}. This robustness makes $Y_{\rm X}$ a reliable and consistent mass proxy that can be used in multiwavelength studies of galaxy clusters.

Studies that have investigated the $L_{\rm X}-Y_{\rm X}$ scaling relation have found deviations from self-similar expectations ($L_{\mathrm{X}}\propto Y_{\rm X}^{3/5}$) \cite[e.g.,][]{Maughan2007, Pratt2009, Eckmiller2011, Lovisari2015, Bahar2022}.

\begin{figure}
    \centering
    \includegraphics[width=\linewidth]{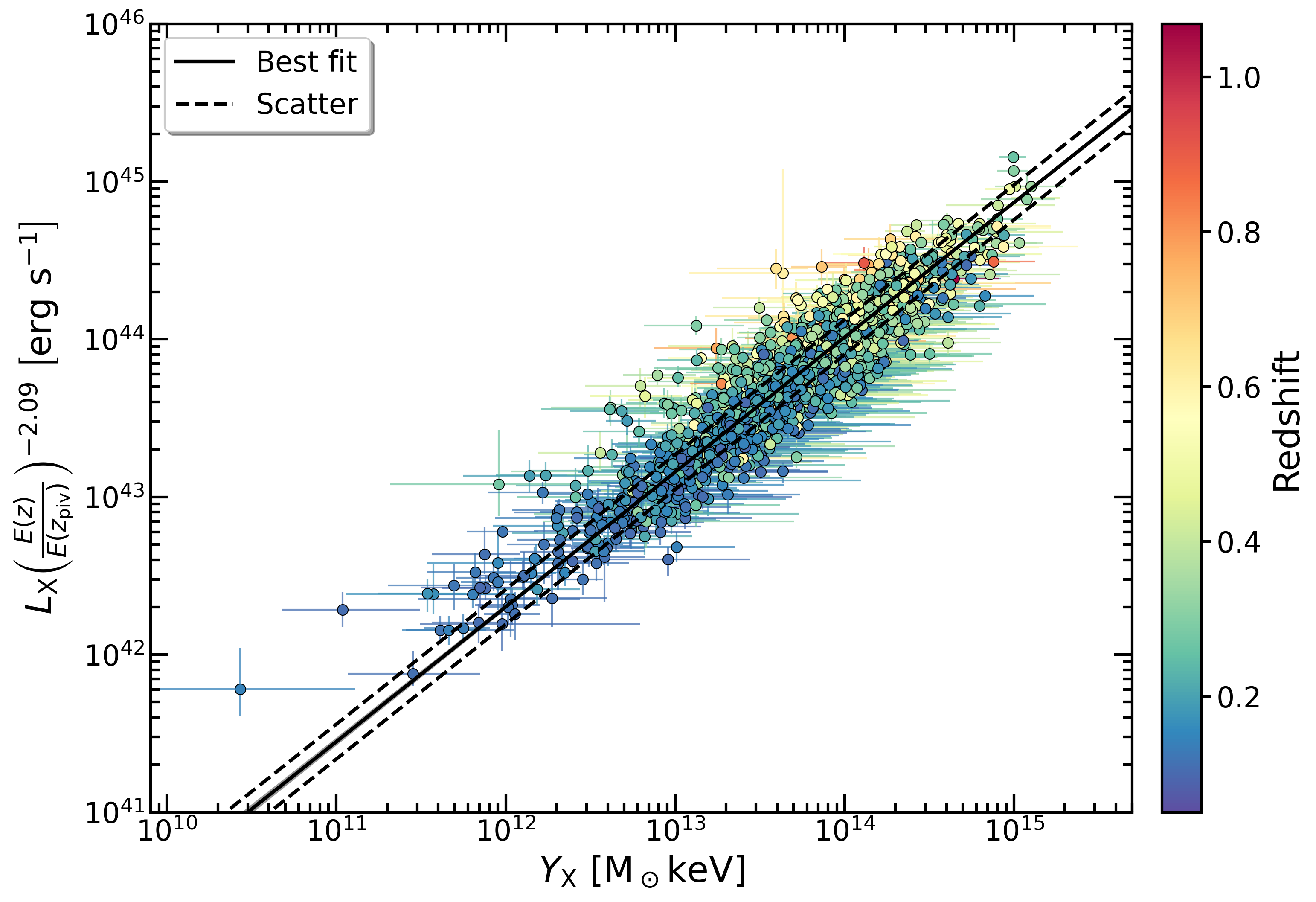}
    \caption{Same as Fig.~\ref{fig:Lx_T_best_fit_model} but for the soft band ($0.5-2.0$~keV energy band) X-ray luminosity--low-scatter mass proxy, $L_{\rm X}-Y_{\rm X}$, scaling relation.}
    \label{fig:Lx_Yx_best_fit_model}
\end{figure}

We found best-fit parameters for the $L_{\rm X}-Y_{\rm X}$ scaling relation of $B=0.855^{+0.009}_{-0.009}$ for the slope, $C=2.09^{+0.16}_{-0.17}$ for the redshift evolution, $A=0.86^{+0.01}_{-0.01}$ for the normalization, and $\sigma_{L_{\rm X}|Y_{\rm X}}=0.25^{+0.01}_{-0.01}$ for the intrinsic scatter. The corresponding model is shown in Fig.~\ref{fig:Lx_Yx_best_fit_model}. Statistically, our best-fit slope is inconsistent with the expected value at the $\sim 28\sigma$ significance level. The redshift evolution parameter also differs from the self-similar value ($C_{\rm SS} = 8/5$), with a tension of about $\sim 3\sigma$. When fixing the redshift evolution to the self-similar value, the slope remains nearly unchanged at $B = 0.866^{+0.009}_{-0.009}$, indicating that our slope result is robust against assumptions on the redshift dependence. Clusters with upper or lower limits on the temperature are assigned $Y_{\rm X}=0$ in the \cite{Bulbul2024} catalog and are therefore omitted from Fig.~\ref{fig:Lx_Yx_best_fit_model}. Their $Y_{\rm X}$ posterior distributions, however, are included in the fitting of the $L_{\rm X}-Y_{\rm X}$ relation.

\begin{figure}
    \centering
    \includegraphics[width=0.8\linewidth]{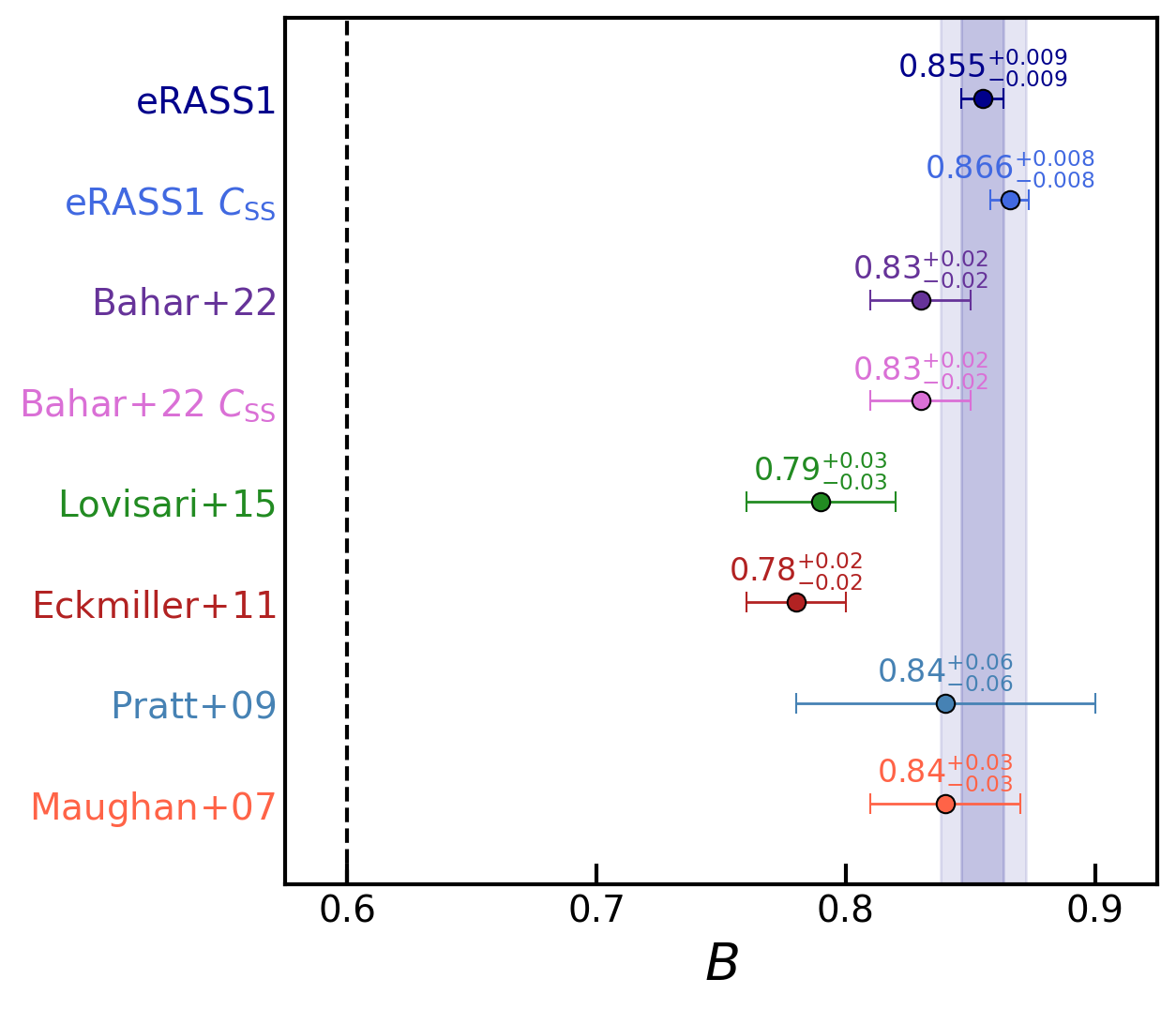}
    \caption{Same as Fig.~\ref{fig:Lx_T_B_comparison} but presenting a comparison of the slope $B$ for the $L_{\rm X}-Y_{\rm X}$ scaling relation, including results from \citet[eFEDS cluster sample]{Bahar2022}, \cite{Lovisari2015}, \cite{Eckmiller2011}, \cite{Pratt2009}, and \cite{Maughan2007}.}
    \label{fig:Lx_Yx_B_comparison}
\end{figure}

These results are compared with previous studies in Fig.~\ref{fig:Lx_Yx_B_comparison} and the lower-left panel of Fig.~\ref{fig:scaling_rel_comparison}. The slope $B$ is in agreement within $1\sigma$ with the findings of \citet[][core excised measurements]{Maughan2007}, \cite{Pratt2009}, and \cite{Bahar2022}. A milder discrepancy is observed when compared to the results of \cite[][$\sim3.4\sigma$]{Eckmiller2011} and \cite[][$\sim2.1\sigma$]{Lovisari2015}. A common feature among all these studies is that the measured slope is consistently steeper than the self-similar prediction. Additionally, our redshift evolution parameter value is higher than that reported by \cite[][$C=1.50^{+0.33}_{-0.35}$]{Bahar2022}, but it is in agreement at the $\sim 1.5\sigma$ level. The normalization measurements are generally consistent across the different studies, with the exception of \cite{Maughan2007}, which reports a lower normalization. This difference is likely due to the use of core-excised X-ray property measurements. Fig.~\ref{fig:Lx_Yx_sigma_comparison} presents a comparison of our measured $\sigma_{L_{\rm X}|Y_{\rm X}}$ with values reported in the literature. Our result is in strong agreement ($1\sigma$) with the studies of \cite{Maughan2007} and \cite{Bahar2022}. Higher values are reported by \cite{Pratt2009}, \cite{Eckmiller2011}, and \cite{Lovisari2015}; however, the latter two do not provide uncertainties for their measurements, making a meaningful statistical comparison with our result not possible.

\subsection{Gas mass--temperature scaling relation}

Another useful scaling relation for probing the total thermal energy of the ICM is the $M_{\rm gas}-T$ relation. As noted earlier, the temperature $T$ reflects the average kinetic energy per particle in the ICM, while $M_{\rm gas}$ is proportional to the total number of particles. Together, these quantities provide a direct measurement of the ICM’s total thermal energy content, making the $M_{\rm gas}-T$ relation a valuable diagnostic for cluster thermodynamics and the impact of nongravitational processes. While $M_{\rm gas}$ and $T$ are linked indirectly, their shared dependence on cluster mass naturally produces a meaningful $M_{\rm gas}-T$ relation, combining the effects of the $M_{\rm gas}-M$ and $M-T$ relations.

\begin{figure}
    \centering
    \includegraphics[width=\linewidth]{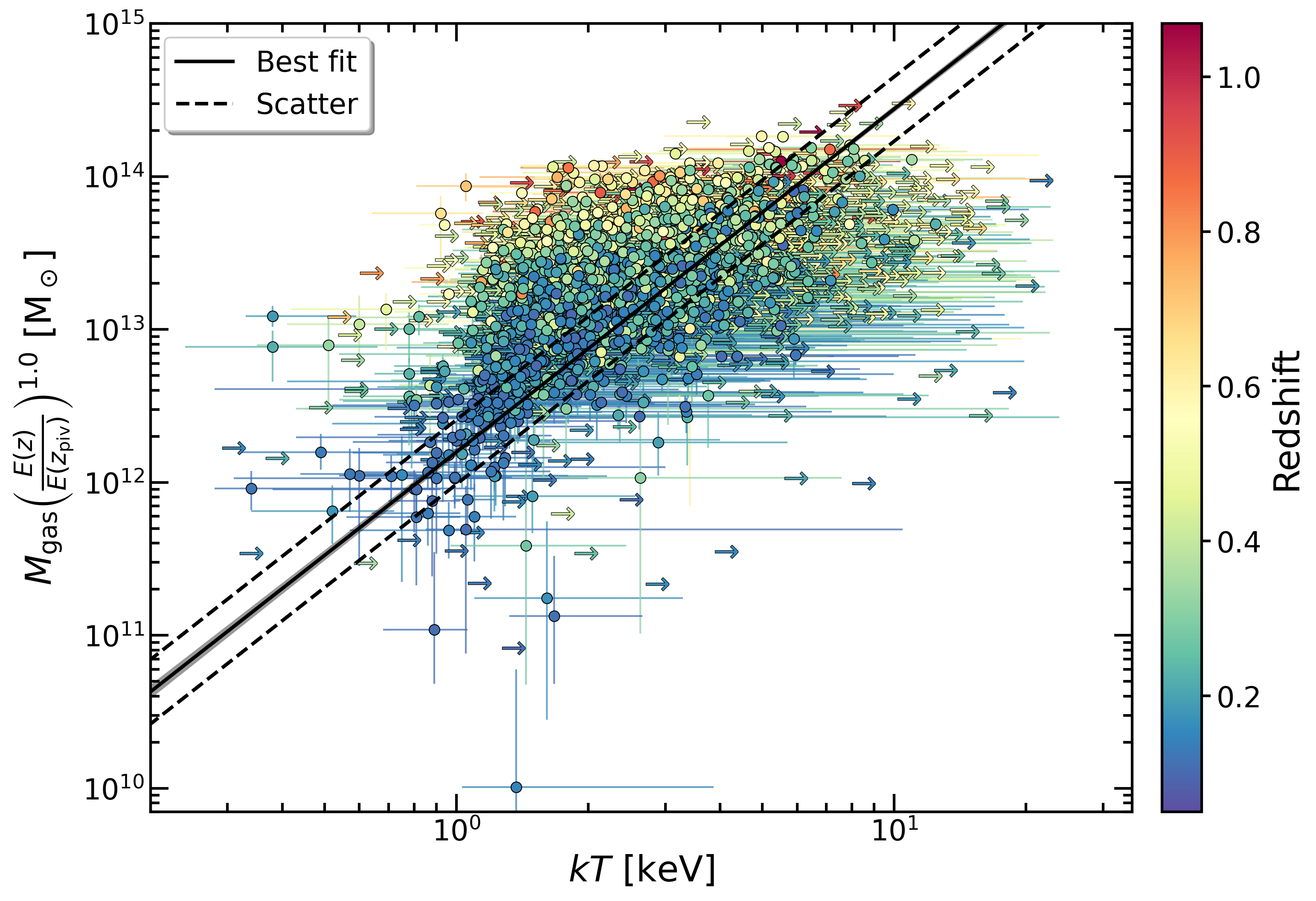}
    \caption{Same as Fig.~\ref{fig:Lx_T_best_fit_model} but for the gas mass--temperature, $M_{\rm gas}-T$, scaling relation.}
    \label{fig:Mgas_T_best_fit_model}
\end{figure}

\begin{figure}
    \centering
    \includegraphics[width=0.8\linewidth]{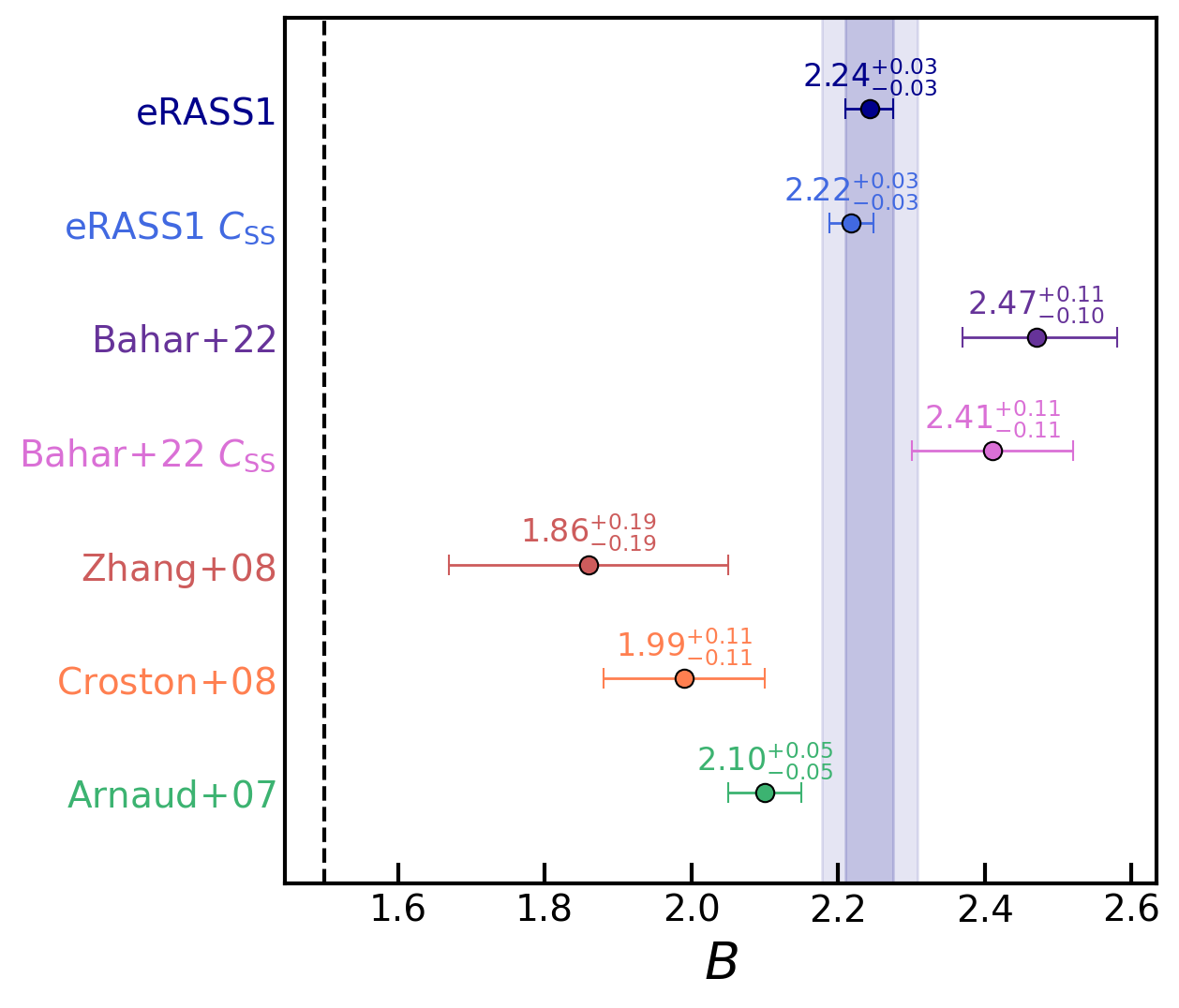}
    \caption{Same as Fig.~\ref{fig:Lx_T_B_comparison} but showing a comparison between the values of the slope $B$ for the $M_{\rm gas}-T$ scaling relation, including results from \citet[eFEDS cluster sample]{Bahar2022}, \cite{Zhang2008}, \cite{Croston2008}, and \cite{Arnaud2007}.}
    \label{fig:Mgas_T_B_comparison}
\end{figure}

Studies of the $M_{\rm gas}-T$ relation have consistently reported a slope of around $2$, significantly steeper than the self-similar expectation \citep[$M_{\rm gas}\propto T^{3/2}$, e.g.,][]{Arnaud2007, Croston2008, Zhang2008, Bahar2022}. This discrepancy suggests that additional physical processes, beyond purely gravitational heating, may play a role in shaping the gas mass–temperature scaling.

As in \cite{Bahar2022}, we also adopted the $L_{\rm X}$-based selection function by converting $M_{\rm gas}$ to $L_{\rm X}$, as no simulation-derived selection function for $M_{\rm gas}$ is available. This one-to-one conversion should not significantly bias our results, since $L_{\rm X}$ and $M_{\rm gas}$ are tightly correlated with relatively low scatter (see Section~\ref{subsec:lx_gmass_relation}).

The resulting best-fit parameters for the $M_{\rm gas}-T$ relation are $B=2.24^{+0.03}_{-0.03}$ for the slope, $C=-1.00^{+0.29}_{-0.30}$ for the redshift evolution, $A=0.44^{+0.02}_{-0.02}$ for the normalization, and $\sigma_{M_{\rm gas}|T}=0.49^{+0.02}_{-0.02}$ for the intrinsic scatter. The corresponding model is presented in Fig.~\ref{fig:Mgas_T_best_fit_model}. Our best-fit $B$ value deviates from the self-similar one by $\sim 25\sigma$. However, we found that the redshift evolution parameter is strongly consistent with the self-similar expectation ($C_{\rm SS}=-1$). Fixing the redshift evolution to the self-similar value yields a slope of $B = 2.22^{+0.03}_{-0.03}$, nearly identical to our baseline result, indicating that the slope is robust to assumptions about redshift dependence. 

Our slope $B$ measurements are compared with earlier studies in Fig.~\ref{fig:Mgas_T_B_comparison}, while the lower-right panel of Fig.~\ref{fig:scaling_rel_comparison} shows the corresponding best-fit relations obtained in this work and in the literature. We found steeper $B$ values than those reported by \citet{Arnaud2007}, \citet{Croston2008}, and \citet{Zhang2008}, but shallower than the results of \citet{Bahar2022}. The difference in $B$ and normalizations appears linked to sample composition: the earlier works focused on massive, nearby clusters, whereas \citet{Bahar2022} included a much larger fraction of low-mass systems, which can strongly influence the fitted slope (see Sect.~\ref{subsec:lx_t_rel}).

Fig.~\ref{fig:Mgas_T_sigma_comparison} presents a comparison of our measured $\sigma_{M_{\rm gas}|T}$ with values reported in the literature. \cite{Zhang2008}, \cite{Croston2008} and \cite{Arnaud2007} report lower values for the intrinsic scatter of the $\sigma_{M_{\rm gas}|T}$ relation, but these measurements are presented without error bars, and therefore a statistical comparison with our findings cannot be made.

\subsection{Comparison with simulations}
The comparison between observed scaling relations and those predicted by hydrodynamical simulations provides a valuable tool for constraining the physical processes governing the ICM and for assessing selection effects in surveys of galaxy groups and clusters. Simulations offer the advantage that individual physical mechanisms, such as galactic winds or AGN feedback, can be included or excluded while keeping the same initial conditions. By comparing the resulting relations with observations, it is therefore possible to identify the processes that most strongly influence the evolution of groups and clusters \citep[see][]{Lovisari2022}. In practice, however, the ability to discriminate between different physical models depends on both the precision of the observations and the intrinsic scatter in the measured scaling relations.

\begin{figure}
    \centering
    \includegraphics[width=\linewidth]{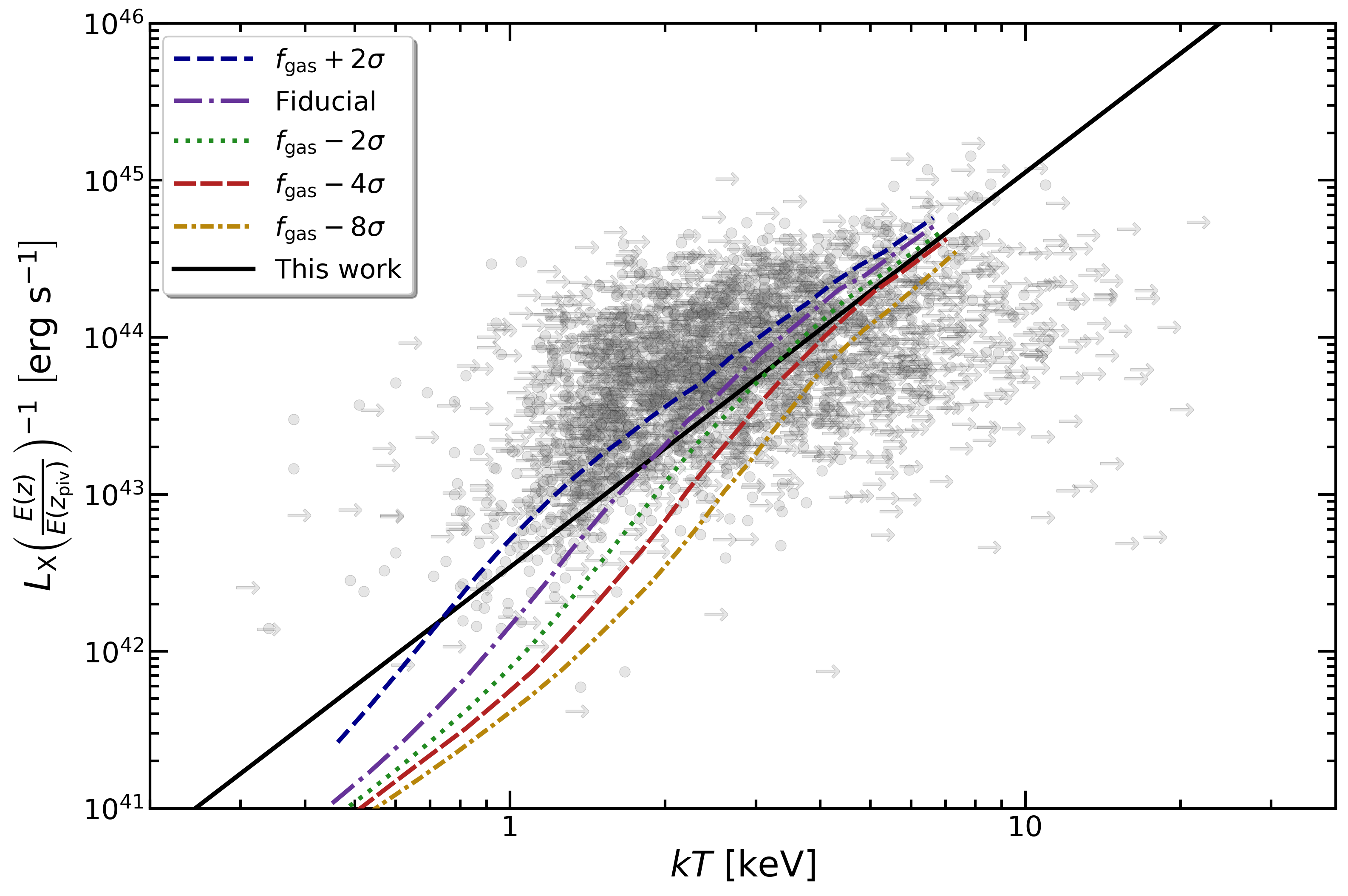}
    \caption{Comparison of the best-fit $L_{\rm X}$–$T$ relation with self-similar redshift evolution derived in this work (solid black line) with the FLAMINGO simulations from \cite{Braspenning2024}. The blue, purple, green, red, and yellow curves correspond to the weak AGN ($f_{\rm gas}+2$), fiducial, strong AGN ($f_{\rm gas}-2$), stronger AGN ($f_{\rm gas}-4$), and strongest AGN ($f_{\rm gas}-8$) feedback models, respectively. }
    \label{fig:Lx_T_simcomparison}
\end{figure}

We compared our results of the $L_{\textrm X}-T$ relation with the predictions of the FLAMINGO simulations \citep[see][for details]{Schaye2023MNRAS, Kugel2023}. The FLAMINGO project is a set of state-of-the-art large cosmological hydrodynamical simulations designed to study the impact of baryonic physics on large-scale structure formation. The simulations include detailed subgrid prescriptions for star formation, stellar evolution, radiative cooling, and feedback from stars and AGN. The subgrid parameters are calibrated using machine-learning-based emulators to reproduce the observed galaxy stellar mass function at $z=0$ and gas fractions in low-redshift groups and clusters. Using the FLAMINGO simulations, \cite{Braspenning2024} studied the evolution of galaxy cluster scaling relations and their profiles, and found that aside from the metallicity of the ICM, the properties of the simulated clusters are in good agreement with those of observed clusters in the real Universe.

Of particular relevance for this work are the different AGN feedback implementations in the FLAMINGO simulations: the fiducial model employs a thermal feedback scheme based on modified Bondi-Hoyle accretion, while alternative simulations explore variations in the feedback strength by shifting the cluster gas fractions relative to observational constraints (e.g.,\ $f_{\rm gas}\pm N\sigma$). In addition, ``jet'' models replace thermal AGN heating with kinetic, jet-driven feedback, allowing the suite to assess the impact of different AGN feedback mechanisms on the thermodynamic properties of galaxy groups and clusters.

Figure~\ref{fig:Lx_T_simcomparison} compares our inferred intrinsic $L_{\rm X}$--$T$ scaling relation with the predictions of the FLAMINGO simulations for different AGN feedback prescriptions \citep{Braspenning2024}, assuming self-similar redshift evolution. Although the intrinsic scatter of our relation ($\sigma_{L_{\rm X}|T}=0.60$) is comparable to the spread among several simulation models, the mean scaling relation is well constrained by the full sample and therefore provides a meaningful comparison with the simulations. Our best-fit relation is inconsistent with the fiducial, weak ($f_{\rm gas}+2\sigma$), and strongest ($f_{\rm gas}-8\sigma$) AGN feedback models. This last result is consistent with the findings of \citet{Eckert2026}, who also found tension with the strongest AGN feedback model when analyzing galaxy groups. For systems with $kT>3$~keV, the strong and stronger AGN feedback models ($f_{\rm gas}-2\sigma$ and $f_{\rm gas}-4\sigma$, respectively) provide the best agreement with the data. At lower temperatures, however, none of the currently available models reproduces our inferred relation. In the low-temperature regime, our sample contains only a few low-luminosity systems ($<10^{42.5}$~erg~s$^{-1}$). In conclusion, the comparison is limited by the intrinsic scatter in the relation, the large fraction of weakly constrained temperature measurements (upper and lower limits), and the relatively small number of low-luminosity systems that probe the low-temperature regime. Consequently, while the present data disfavor some feedback prescriptions, they do not yet provide sufficient discriminatory power to uniquely identify the preferred model.


%
\section{Discussion}
\label{sec:discussion}

The large sample of galaxy groups and clusters analyzed in this study enables tight constraints on the parameters describing the scaling relations, for example, the slope and the redshift evolution parameter. The resulting uncertainties are significantly smaller than those reported in previous studies, reflecting the statistical power of the eRASS1 galaxy cluster and group sample. In the next sections, we discuss our analysis results in more detail.

\subsection{Deviations from self-similar predictions}

The slopes, $B$, of the scaling relations between X-ray observables analyzed in this work exhibit statistically significant deviations from the predictions of the self-similar model, with departures exceeding $25\sigma$. Our results agree with previous observational studies that also find a departure from the self-similar scenario \cite[e.g.,][]{Maughan2007, Eckmiller2011, Zhang2011, Kettula2015, Lovisari2015, Mantz2016, Migkas2020, Bahar2022}. These slope discrepancies are commonly attributed to nongravitational processes in the ICM, such as AGN feedback, galactic winds, and star formation \citep[e.g.,][]{Voit2005}. Such processes inject additional energy into the ICM, altering its thermodynamic structure by heating the gas and raising its entropy, particularly in the central regions of clusters. Such mechanisms disrupt the simplifying assumptions of self-similarity, including spherical symmetry and purely gravitational heating. Theoretical studies and hydrodynamical simulations further support the breakdown of self-similarity due to these nongravitational effects \cite[e.g.,][]{Babul2002, Truong2018, Henden2018, Henden2019}.

\subsection{Galaxy groups in the sample}
\label{subsec:galgroups_discussion}

As discussed in Section~\ref{subsec:lx_t_rel}, one of our main results is the discrepancy in the slope values of the $L_{\mathrm{X}}-T$ and $L_{\mathrm{X}}-M_{\rm gas}$ relations when compared to the findings of \citet[][eFEDS sample]{Bahar2022}. It has been shown that galaxy groups tend to follow a steeper $L_{\mathrm{X}}-T$ scaling relation than clusters, owing to larger scatter in their ICM properties \citep[e.g.,][]{Maughan2012, Osmond2004, Lovisari2021}. In the full eRASS1 cluster catalog \citep{Bulbul2024}, the fraction of galaxy groups ($<10^{14}$~M$_\odot$) is approximately $23\%$, indicating that the eRASS1 survey's shallowness favors the detection of higher-mass systems. Even after decreasing the extent likelihood threshold (see Section~\ref{subsec:sample_selection}) to enlarge our sample, the proportion of low-mass systems remains limited. For instance, selecting \extlike$>7$ increases the sample size by $70\%$ and raises the group fraction to $13\%$, but also increases the sample contamination to $5\%$. In this analysis, we prioritized maintaining low contamination to ensure the reliability of our sample and the robustness of our best-fit models. Future analysis of deeper eRASS:5 observations will improve sensitivity to low-mass systems, allowing for a better assessment of their influence on scaling relations, for example, by allowing a broken power-law fit to the entire galaxy group and cluster sample. 

A potential source of systematic uncertainty in the group results is the assumed metallicity of $0.3~Z_{\odot}$. If the true metallicities are higher, the inferred temperatures and luminosities could be biased low, whereas lower metallicities could bias them high. However, the magnitude of this effect is expected to be small compared to the statistical uncertainties of our measurements. For example, the median $kT$ of our sample is $2.27$~keV. For a $2.27$~keV object, changing metallicity from $0.3$ to $0.4~Z_{\odot}$ only results in a $\sim 5\%$ change in luminosity. Moreover, other mechanisms in lower-mass clusters and groups, primarily line emission from collisionally excited ions, make increasingly important contributions to the overall X-ray emission \citep{Lovisari2021, Lovisari2022}. Given the limited nature of our X-ray data, it is not possible to determine the impact of such a mechanism on our sample; therefore, it is referred to future work.

\subsection{Cool-core effects}
\label{subsec:cc_discussion}

Evidence indicates that the most massive, relaxed clusters follow a self-similar $L_{\mathrm{X}}-T$ scaling relation when their cores are excluded \citep{Maughan2012}, as nongravitational effects such as cooling and AGN feedback are minimal in the outer regions. However, we are unable to test this with our current sample for two main reasons. First, eRASS1 clusters have low photon counts, and excising the core further reduces photon statistics, often rendering X-ray measurements unreliable or unfeasible. Second, a separate selection function based on core-excised observables would be required, but this in turn relies on accurate PSF modeling that is not yet available in the simulations.

\subsection{$R_{500}$ estimations}

In this work, temperatures are less affected by the choice of $R_{500}$ because they represent an average quantity (see Sect.~\ref{subsec:data_analysis}) rather than an integrated one. Consequently, the temperature depends less strongly on the precise value of $R_{500}$ than integrated quantities such as luminosity and gas mass.

The estimation of $M_{500}$, and therefore $R_{500}$, was performed within the eRASS1 cosmology pipeline \citep{Ghirardini2024}, where cluster masses are inferred by constructing the full probability density function of mass conditioned on the observed count rate and redshift. The adopted $R_{500}$ values are based on weak-lensing calibrated masses and therefore do not rely on assumptions about hydrostatic equilibrium or nonthermal pressure support. The approach combined the count-rate--mass scaling relation calibrated with weak-lensing shear measurements with the survey selection function, measurement uncertainties, and the cosmological mass function, so that uncertainties in the scaling relation parameters, including its slope, are propagated into the mass estimates \citep[see][for details]{Bulbul2024}. The method has been validated using mock observations, showing that the recovered mass PDFs reproduce the underlying cluster mass distribution without introducing systematic bias. Consequently, the inferred $R_{500}$ values already account for these uncertainties, which are propagated to derived quantities such as $M_{\rm gas}$ and $L_{\rm X}$. We therefore expect that the impact of uncertainties in the count-rate-mass relation on the measured scaling relations is properly captured within the current framework.

\subsection{Selection biases}

Selection biases, such as the Malmquist bias \citep{Malmquist1922}, which arises because brighter sources are detectable at greater distances, are a common concern in studies of astrophysical samples, especially flux-limited samples. In our analysis, these effects are carefully modeled and included through the selection function. The determination of the selection function for the eRASS1 sample \citep{Clerc2024} builds upon and improves the method used in the eFEDS analysis \citep{Bahar2022}. The selection function has been carefully modeled and tested on end-to-end simulations of the eROSITA sky, which include a realistic, representative distribution of galaxy groups and clusters \citep{Comparat2020, Seppi2022}. Its reliability has been demonstrated in recent cosmological constraints from eROSITA cluster abundances \citep{Ghirardini2024}. We therefore conclude that the observed departures from self-similar scaling in our results cannot be attributed to selection biases.

There is evidence that relaxed, cool-core clusters exhibit less scatter in scaling relations than mergers \citep{Vikhlinin2009, Mantz2010}, and reducing scatter is crucial for tightening cosmological constraints from cluster counts. \cite{Sanders2025} recently derived the intrinsic distribution of eRASS1 morphological parameters within a Bayesian framework, constructed their scaling relations with redshift and luminosity, and examined their distributions in redshift and luminosity bins. These results provide a basis for future analyses that employ morphology-based selection functions and dynamical–state–dependent mass functions to study X-ray scaling relations among observables.

\subsection{Temperature offsets}

Previous studies comparing eROSITA with XMM-{\it Newton} and {\it Chandra} have found inconsistent temperature offsets, with results ranging from eROSITA measuring lower ($>25\%$) to slightly higher temperatures depending on the studied cluster, chosen energy band, and observation mode, pointed observation or survey \citep{Sanders2022, Whelan2022, veronica2022, Liu2023}. These works relied on single-cluster analyses, which are prone to noise and methodological differences, including variations in metal abundances and Galactic absorption assumptions. \cite{Turner2022} analyzed eight eFEDS clusters and found eROSITA temperatures $25\pm9\%$ lower than XMM-{\it Newton} in the broad band, though different energy bands were used. \cite{Migkas2024} recently compared eROSITA galaxy cluster temperatures with those from {\it Chandra} and XMM-{\it Newton} using a large sample. They measured 186 overlapping regions ($<R_{500}$) with {\it Chandra} and 71 with XMM-{\it Newton} in the full ($0.7-7.0$~keV), soft ($0.5-4.0$~keV), and hard ($1.5-7$~keV) X-ray energy bands. eROSITA temperatures were consistently lower, by $15-38\%$ depending on cluster temperature and energy band, with the largest differences for hot clusters, smaller offsets for cool systems, and the strongest mismatch in the hard band. A broken power-law fit showed a break in the scaling relation around $1.7-2.7$~keV, depending on the energy band. These results may suggest that our best-fit $L_{\mathrm{X}}-T$ and $M_{\rm gas}-T$ relations are steeper than the self-similar expectation because hotter systems are, in reality, cooler. Nonetheless, our findings agree with previous scaling relation studies, mostly based on {\it Chandra} and XMM-{\it Newton} data, which also report steeper $L_{\mathrm{X}}-T$ and $M_{\rm gas}-T$ relations than predicted by the self-similar model \citep[e.g.,][]{Arnaud2007, Pratt2009, Eckmiller2011, Kettula2015, Lovisari2015, Giles2016, Migkas2020}. 

In Sect.~7.9 of \cite{Migkas2024}, the authors report an agreement between the eRASS1 temperatures \citep{Bulbul2024} and their spectroscopically derived temperatures using full-band data, providing the associated conversion relations \citep[see Table~A.3 in][]{Migkas2024}. We applied the eROSITA–XMM-{\it Newton} temperature relation from \citet{Migkas2024}, converting our temperature chains and fitting the $L_{\rm X}-T$ relation. We obtained a slope of $B=2.58^{+0.04}_{-0.04}$, a redshift evolution parameter of $C=0.93^{+0.55}_{-0.56}$, an intrinsic scatter of $\sigma_{L_{\rm X}|T}=0.67^{+0.04}_{-0.04}$, and a normalization of $A=0.44^{+0.04}_{-0.04}$. The slope is consistent with our results in Section~\ref{subsec:lx_t_rel}, while the redshift evolution is lower, but still compatible within uncertainties. The intrinsic scatter increases, attributable to the additional uncertainties introduced by the temperature offset, in addition to the existing measurement errors. Fixing the redshift evolution to the self-similar value, $C=1$, yields results that are fully consistent with those obtained when $C$ is treated as a free parameter. We therefore did not find evidence that the inferred normalization depends on whether the redshift evolution is fixed or fitted.

\subsection{Redshift evolution}

Understanding the scaling relation redshift evolution is essential for cosmological uses of clusters. However, this has been a challenging task due to the lack of high-quality data for large, well-characterized samples of high-redshift clusters, and the need for a consistent low-redshift counterpart. Differences in sample selection, analysis, or instrumentation between local and distant data can hide or mimic a plausible evolution. \cite{Giodini2013} and \cite{Lovisari2022} note that there is still no clear consensus in the literature regarding the redshift evolution of scaling relations, despite numerous past studies. The sample of \citet{Bahar2022}, covering $0.017<z<0.940$, shows self-similar redshift evolution for all examined scaling relations (the same as those in this work) at a significance below $2.5\sigma$. Similarly, we found only mild evolution, consistent with self-similar predictions at the $1-3\sigma$ confidence level.

Our results show no significant deviation from self-similar expectations in the evolution of galaxy cluster scaling relations out to $z\sim1.2$, suggesting that clusters formed or reached equilibrium relatively recently compared to the epoch of observation. However, uncertainties remain, and both weaker and stronger departures from self-similar predictions cannot yet be ruled out.

\subsection{Intrinsic scatter}

As mentioned in \cite{Lovisari2021}, scaling relations deviate from self-similar predictions partly due to their intrinsic scatter. In relations involving $L_{\rm X}$, the scatter can be as large as $50-70$\% at a fixed $T$ (see Fig.~\ref{fig:Lx_T_sigma_comparison}). Our results for this relation give an intrinsic scatter of $\sim57\%$. Previous studies \citep[e.g.,][]{Maughan2007} have shown that this large variation is mainly due to the cluster cores, where cool-core systems show higher luminosities. Excluding the central $0.15R_{500}$, for example, reduces the intrinsic scatter to $10-20$\%. However, excluding the central part of eRASS1 systems is not possible with current data (see Section~\ref{subsec:cc_discussion}).

For the other scaling relations presented in this work, however, the situation is less clear. Reported scatter values in the literature span a broad range, often depending on sample selection, measurement methods, and other factors. Some studies do not report errors on their scatter, while others report substantially larger scatter, reflecting both astrophysical diversity among clusters and systematic differences in analysis. As a result, a direct comparison with our findings is not straightforward.

\subsection{Cosmological assumptions}

Given the high statistical precision of our results (e.g., uncertainties of $<2\%$ on some scaling-relation slopes), it is important to consider the potential impact of uncertainties in the adopted cosmological parameters. Variations in quantities such as the Hubble constant can affect the redshift distribution and, consequently, the derived slopes and normalization. However, a full assessment would require multiple realizations of the digital twin and corresponding selection functions, a computationally prohibitive task beyond the scope of this work.

In this context, the halo mass function constitutes another potential source of systematic uncertainty. Its calibration is typically derived from numerical simulations, and over the past decades, several alternative formulations have been proposed for use in cosmological analyses \citep[see review by][]{Asgari2023}. To quantify the impact of this choice on our results, we tested another representative mass function model by \cite{Despali2016}, and assess its effects (through $P(X|z)$, see Sect.~\ref{sec:like_fitting}) on the $L_{\rm X}-T$ scaling relation. From this test we obtained: a slope of $B=2.50^{+0.04}_{-0.04}$, a redshift evolution parameter of $C=1.91^{+0.37}_{-0.36}$, a normalization of $A=0.42^{+0.02}_{-0.02}$, and an intrinsic scatter of $\sigma_{L_{\rm X}|T}=0.57^{+0.03}_{-0.03}$. These results are consistent with the results shown in Sect.~\ref{subsec:lx_t_rel}. This test demonstrates that the statistical uncertainty on the derived slopes and normalization of the scaling relation remains similar even after using a different halo mass function.

\subsection{Systematic uncertainties}

The uncertainties quoted throughout this work represent the statistical uncertainties of our fit, including the intrinsic scatter of the adopted mass--observable scaling relations. They do not include systematic uncertainties associated with calibrating the external scaling relations themselves (see Sect.~\ref{sec:like_fitting}). As a robustness test, we repeated the analysis while marginalizing over the normalization of the \cite{Chiu2022} $M_{\rm gas}-M$ relation. This leads to best-fitting parameters of $B=1.304^{+0.008}_{-0.008}$, a redshift evolution parameter of $C=2.33^{+0.07}_{-0.07}$, a normalization of $A=1.006^{+0.006}_{-0.006}$, and an intrinsic scatter of $\sigma_{L_{\rm X}|T}=0.208^{+0.006}_{-0.005}$. Compared with the values reported in Table~\ref{tab:best_fit_parameters}, the changes in the best-fitting parameters are substantially larger than their statistical uncertainties, while the statistical uncertainties themselves remain essentially unchanged. In particular, the shift in the slope $B$ corresponds to approximately $7\sigma$, illustrating that the high statistical precision of our measurements makes the inferred scaling relation parameters sensitive to relatively small changes in the external mass calibration. The evolution parameter $C$ also changes with the adopted calibration, although the corresponding shift is less than $2\sigma$. This demonstrates that uncertainties in the external mass calibration can constitute an important source of systematic uncertainty.

We did not propagate the uncertainty in the slope of the \cite{Chiu2022} relation, as its explicit redshift dependence makes it difficult to disentangle the effect of the calibration uncertainty from the intrinsic redshift evolution of the scaling relation. A full treatment of these calibration uncertainties and their covariance is beyond the scope of the present work.


%
\section{Summary}
\label{sec:summary}
We used the largest ICM- and IGrM-detected sample of galaxy groups and clusters to date from the eRASS1 catalog to investigate the X-ray scaling relations $L_{\mathrm{X}}$–$T$, $L_{\mathrm{X}}$–$M_{\rm gas}$, $L_{\mathrm{X}}$–$Y_{\rm X}$, and $M_{\rm gas}$–$T$. The sample comprises $3061$ objects spanning $0.05<z<1.07$ and $1.1\times10^{13}<M_{500}/$M$_{\odot}<1.6\times10^{15}$. Our analysis includes a detailed treatment of the cluster selection function. We fit the scaling relations, allowing the normalization, slope, redshift evolution, and intrinsic scatter in both $X$ and $Y$ to vary freely. In a second run, the redshift evolution exponent was fixed to the corresponding self-similar value. Our results are summarized as follows:
\begin{itemize}
    \item The slopes of the X-ray scaling relations studied here deviate significantly from self-similar expectations, but are consistent with previously reported observational and simulation results. Our findings support the interpretation that these departures from self-similarity arise from nongravitational processes, such as feedback mechanisms.
    \item With the large sample of groups and clusters analyzed here, we robustly confirmed that the redshift evolution of the X-ray scaling relations remains broadly consistent with the self-similar model out to $z\sim 1.2$. When the evolution exponent is fixed to its self-similar value, the best-fit slopes change by less than $1\sigma$ relative to the free-exponent case.
    \item The intrinsic scatter of the $L_{\rm X}-T$ relation is consistent with values reported in previous studies, confirming agreement with earlier observational constraints. For the other scaling relations, however, the wide range of scatter values found in the literature makes direct comparison less straightforward. In both cases, the level of scatter remains similar whether the redshift evolution parameter is left free or fixed to the self-similar value, suggesting that our inference of scatter is robust to assumptions about the redshift dependence. 
    \item Compared to the eFEDS sample analyzed in \citet{Bahar2022}, our galaxy group and cluster sample contains relatively fewer low-mass systems, which may account for the discrepancy in slope values. Future, larger samples of groups and clusters from deeper data will provide a clearer understanding of the impact of galaxy groups on scaling relation fits.
    \item The comparison of our inferred $L_{\rm X}$--$T$ scaling relation with the FLAMINGO simulations disfavors the fiducial, weak, and strongest feedback models, while the strong and stronger feedback models provide the best agreement for systems with $kT>3$~keV. At lower temperatures, none of the currently available models reproduces the inferred relation. In general, this comparison remains limited by the intrinsic scatter of the relation, the large fraction of systems with upper and lower temperature limits, and the relatively small number of low-luminosity systems in the low-temperature regime. Consequently, the present sample can reject some feedback prescriptions but does not yet uniquely identify the preferred AGN feedback model.
\end{itemize}

Even with shallow observations, eROSITA has successfully imaged hundreds of galaxy clusters and groups and enabled robust measurements of their X-ray properties, even from only a few photons. Our work demonstrates that those results can place tight constraints on the X-ray scaling relations of clusters and groups. With future, deeper eROSITA catalogs, the significantly increased number of detected systems, particularly at the group scale, will allow us to extend these constraints to lower masses and higher redshifts. This will provide a unique opportunity to probe the interplay between gravitational and nongravitational processes in structure formation, thereby deepening our understanding of the physics governing the evolution of galaxy groups and clusters.


%
\begin{acknowledgement}
The authors thank the referee for their helpful and constructive comments, which have improved the manuscript. We are also grateful to J. Braspenning for providing the $L_{\rm X}-T$ fitting results from the FLAMINGO simulations.
\\
This work is based on data from \rosi, the soft X-ray instrument aboard SRG, a joint Russian-German science mission supported by the Russian Space Agency (Roskosmos), in the interests of the Russian Academy of Sciences represented by its Space Research Institute (IKI), and the Deutsches Zentrum f{\"{u}}r Luft und Raumfahrt (DLR). The SRG spacecraft was built by the Lavochkin Association (NPOL) and its subcontractors and is operated by NPOL with support from the Max Planck Institute for Extraterrestrial Physics (MPE). The development and construction of the \rosi\ X-ray instrument were led by MPE, with contributions from the Dr. Karl Remeis Observatory Bamberg \& ECAP (FAU Erlangen-Nuernberg), the University of Hamburg Observatory, the Leibniz Institute for Astrophysics Potsdam (AIP), and the Institute for Astronomy and Astrophysics of the University of T{\"{u}}bingen, with the support of DLR and the Max Planck Society. The Argelander Institute for Astronomy of the University of Bonn and the Ludwig Maximilians Universit{\"{a}}t Munich also participated in the science preparation for \rosi.

\\

The eROSITA data shown here were processed using the eSASS/NRTA software system developed by the German eROSITA consortium.

\\

E. Bulbul, S. Zelmer, X. Zhang, and E. Artis acknowledge financial support from the European Research Council (ERC) Consolidator Grant under the European Union’s Horizon 2020 research and innovation program (grant agreement CoG DarkQuest No 101002585). N. Clerc was financially supported by CNES. 
A. Liu acknowledges the support from the National Natural Science Foundation of China (Grant No. 12588202). A. Liu is supported by the China Manned Space Program with grant no. CMS-CSST-2025-A04.

This work made use of SciPy \citep{jones_scipy_2001}, matplotlib, a Python library for publication-quality graphics \citep{Hunter2007}, Astropy, a community-developed core Python package for Astronomy \citep{Astropy2013}, NumPy \citep{van2011numpy}, Colossus \citep{Diemer2018}, and ChainConsumer \citep{Hinton2016}.

\end{acknowledgement}


\bibliography{eRASS1Scl}   


\begin{appendix}


\section{Parameter constraints}
\label{app:parameter_constraints}

\begin{figure}
    \centering
    \includegraphics[width=\columnwidth]{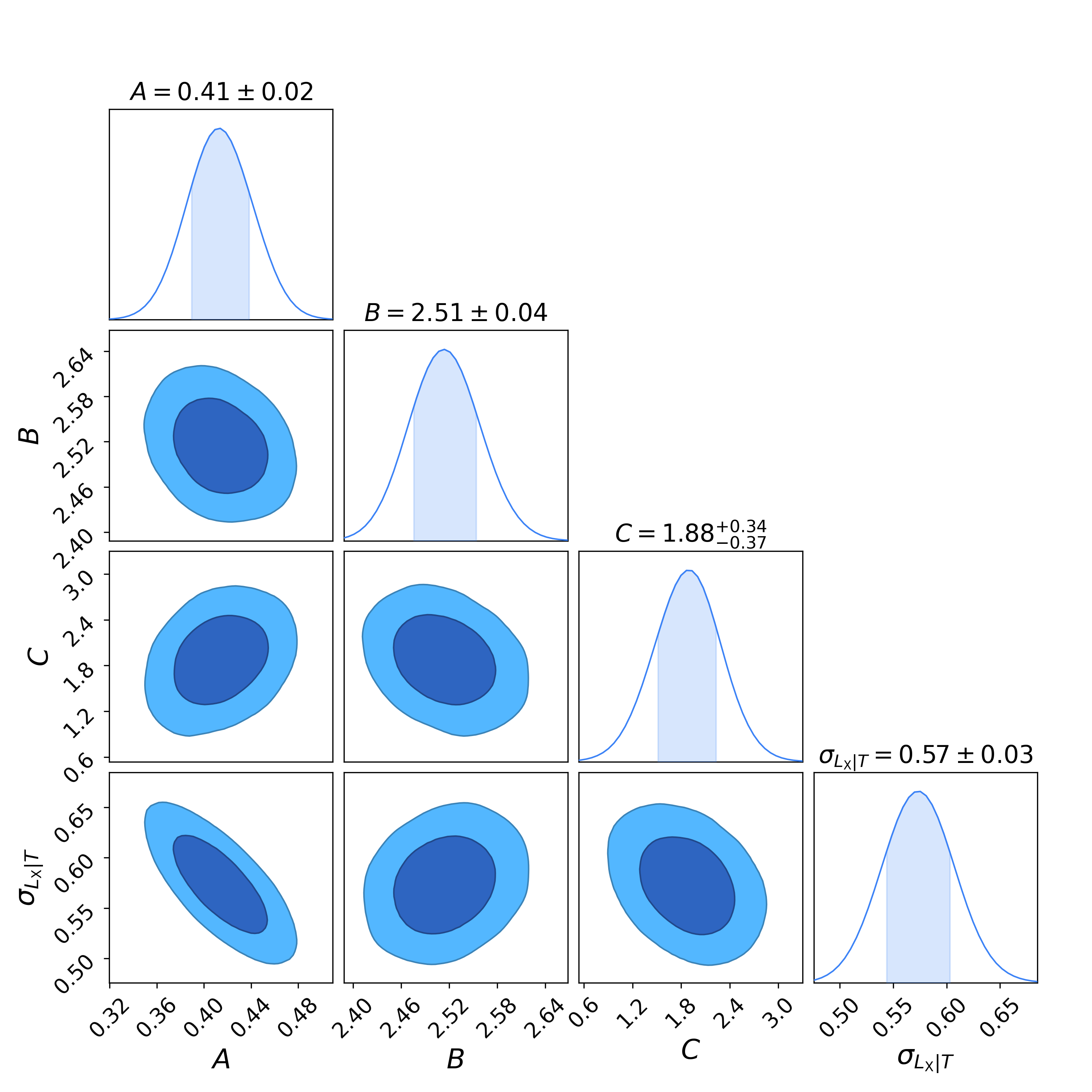}
    \caption{Parameter constraints of the $L_{\rm X}-T$ scaling relation. The off-diagonal panels display the joint posteriors, with contours representing the 68\% (dark blue) and 95\% (light blue) confidence levels. The diagonal panels show the marginalized posterior distributions, and the shaded light blue areas show the $1\sigma$ region.}
    \label{fig:Lx_T_relation_cornerplot}
\end{figure}

\begin{figure}
    \centering
    \includegraphics[width=\columnwidth]{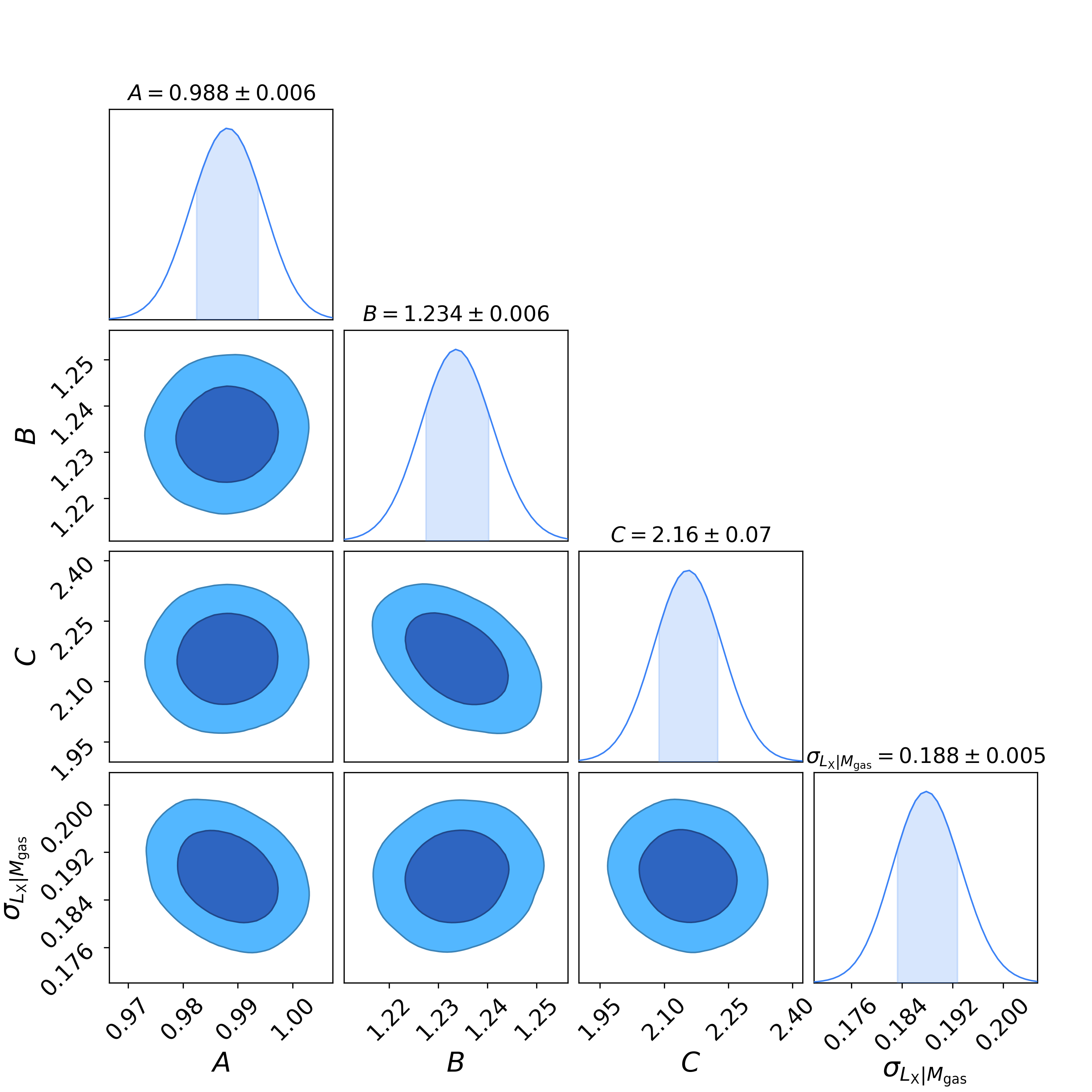}
    \caption{Same as Fig.~\ref{fig:Lx_T_relation_cornerplot}, but for the $L_{\rm X}-M_{\rm gas}$ scaling relation.}
    \label{fig:Lx_Mgas_relation_cornerplot}
\end{figure}

Corner plots displaying the parameter distributions from the MCMC chains are shown in Figs.~\ref{fig:Lx_T_relation_cornerplot}, \ref{fig:Lx_Mgas_relation_cornerplot}, and \ref{fig:Lx_Yx_relation_cornerplot}, for the $L_{\rm X}-T$, $L_{\rm X}-M_{\rm gas}$ and $L_{\rm X}-Y_{\rm X}$ scaling relation analysis, respectively. These plots show the marginalized and joint posterior distributions of the model parameters for the scaling relation fittings.

\begin{figure}
    \centering
    \includegraphics[width=\columnwidth]{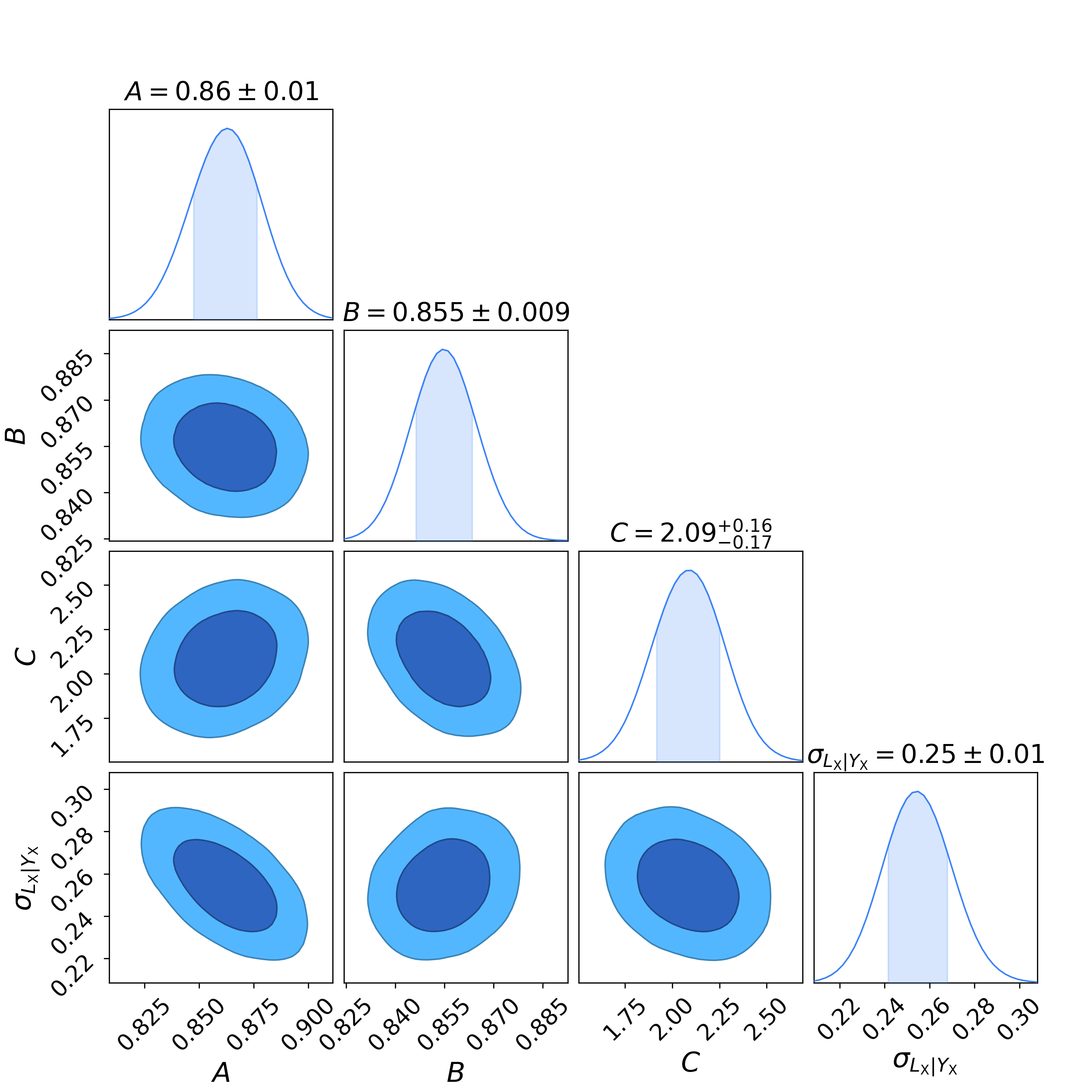}
    \caption{Same as Fig.~\ref{fig:Lx_T_relation_cornerplot} but for the $L_{\rm X}-Y_{\rm X}$ scaling relation.}
    \label{fig:Lx_Yx_relation_cornerplot}
\end{figure}

\begin{figure}
    \centering
    \includegraphics[width=\columnwidth]{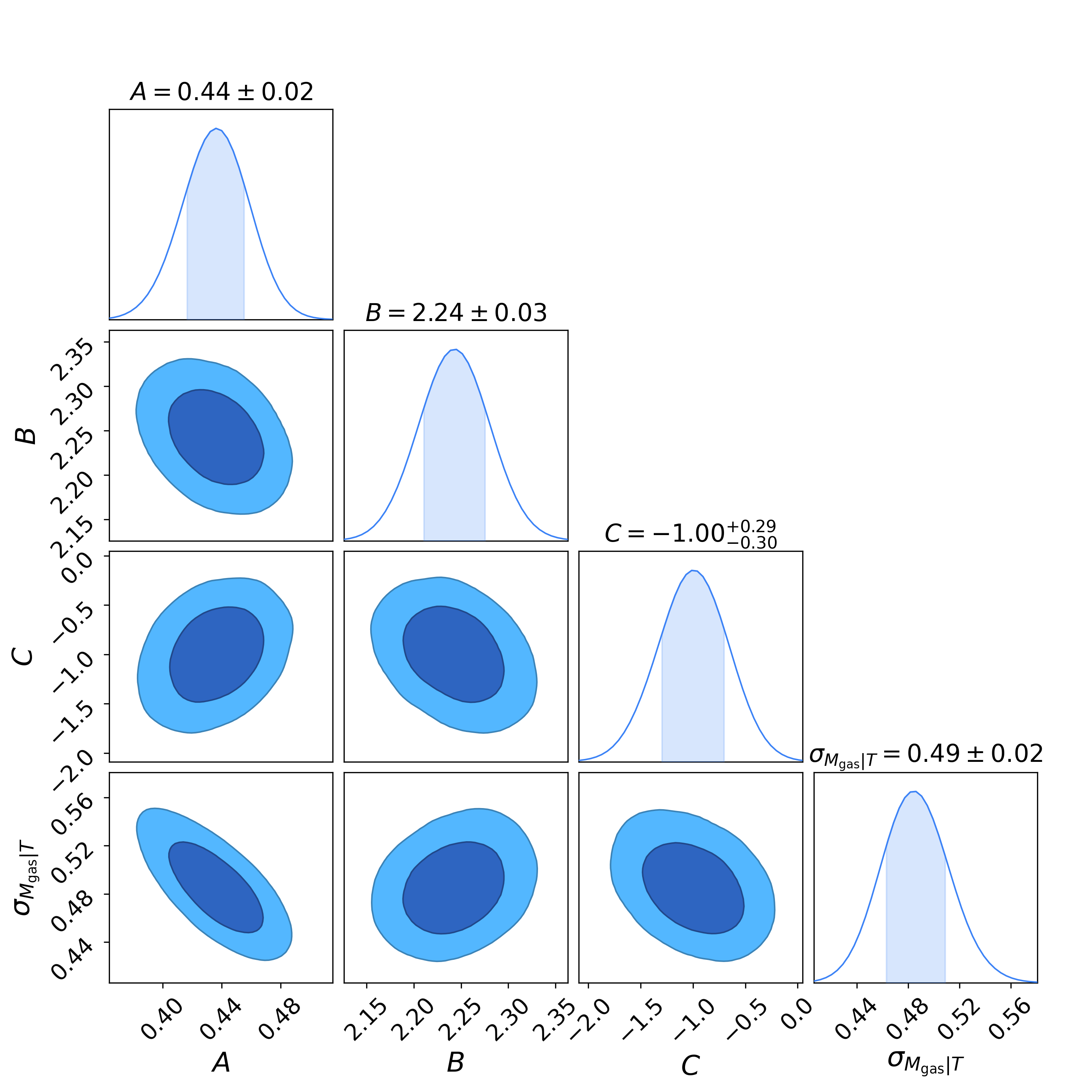}
    \caption{Same as Fig.~\ref{fig:Lx_T_relation_cornerplot} but for the $M_{\rm gas}-T$ scaling relation.}
    \label{fig:Lx_Yx_relation_cornerplot}
\end{figure}


\section{Scatter comparison}
\label{app:scatter_comparison}

\begin{figure}
    \centering
    \includegraphics[width=0.8\linewidth]{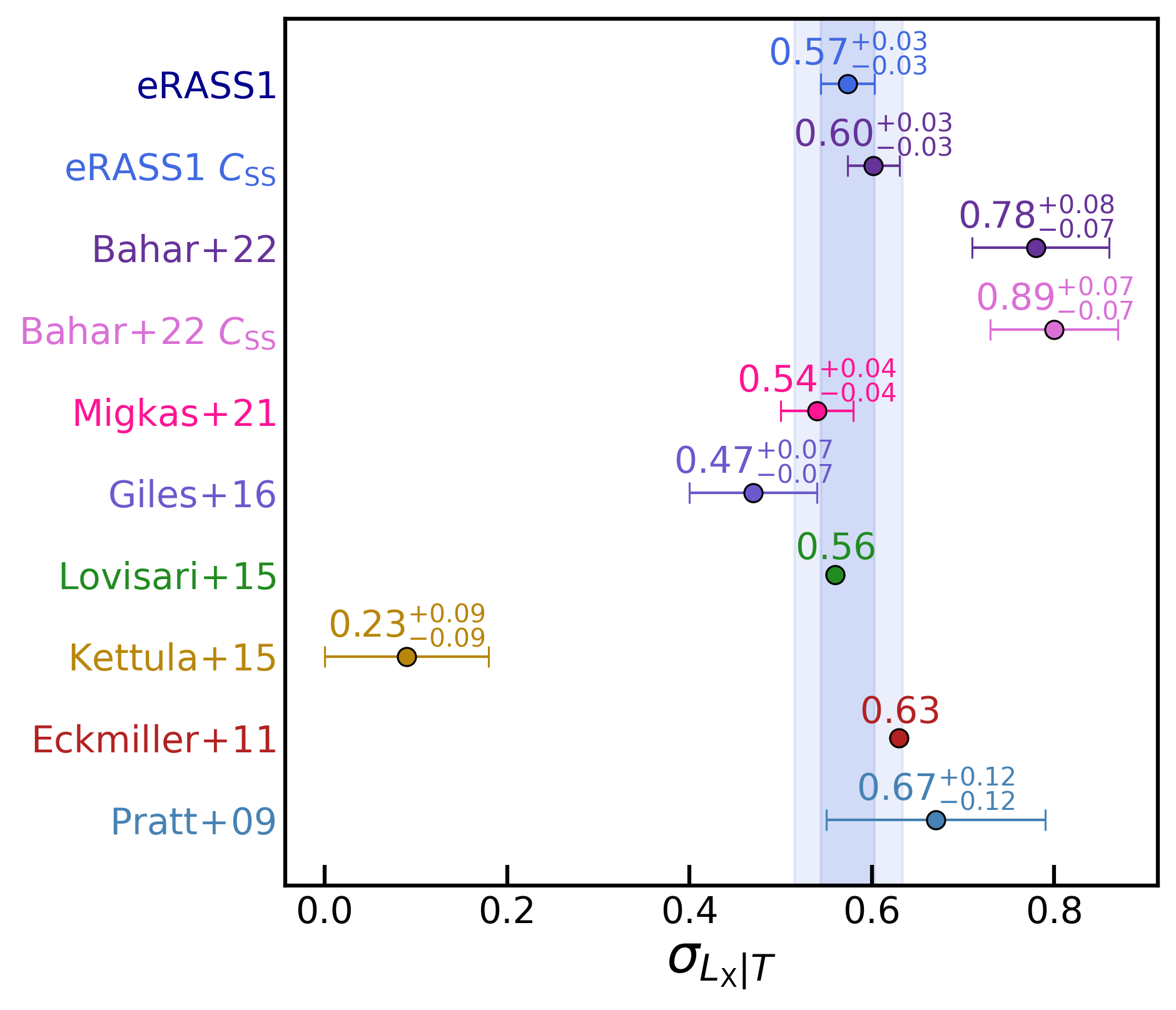}
    \caption{Comparison of the intrinsic scatter $\sigma_{L_{\rm X}|T}$ for the $L_{\rm X}-T$ scaling relation with literature results from \citet[eFEDS cluster sample]{Bahar2022}, \citet[XXL cluster sample]{Giles2016}, \citet[eeHIFLUGCS]{Migkas2021}, \cite{Lovisari2015}, \cite{Kettula2015}, \cite{Eckmiller2011}, and \cite{Pratt2009}. All the shown errors are at the $1\sigma$ level. The vertical shaded areas represent our best-fit measurements at the $1-2\sigma$ confidence level.}
    \label{fig:Lx_T_sigma_comparison}
\end{figure}

\begin{figure}
    \centering
    \includegraphics[width=0.8\linewidth]{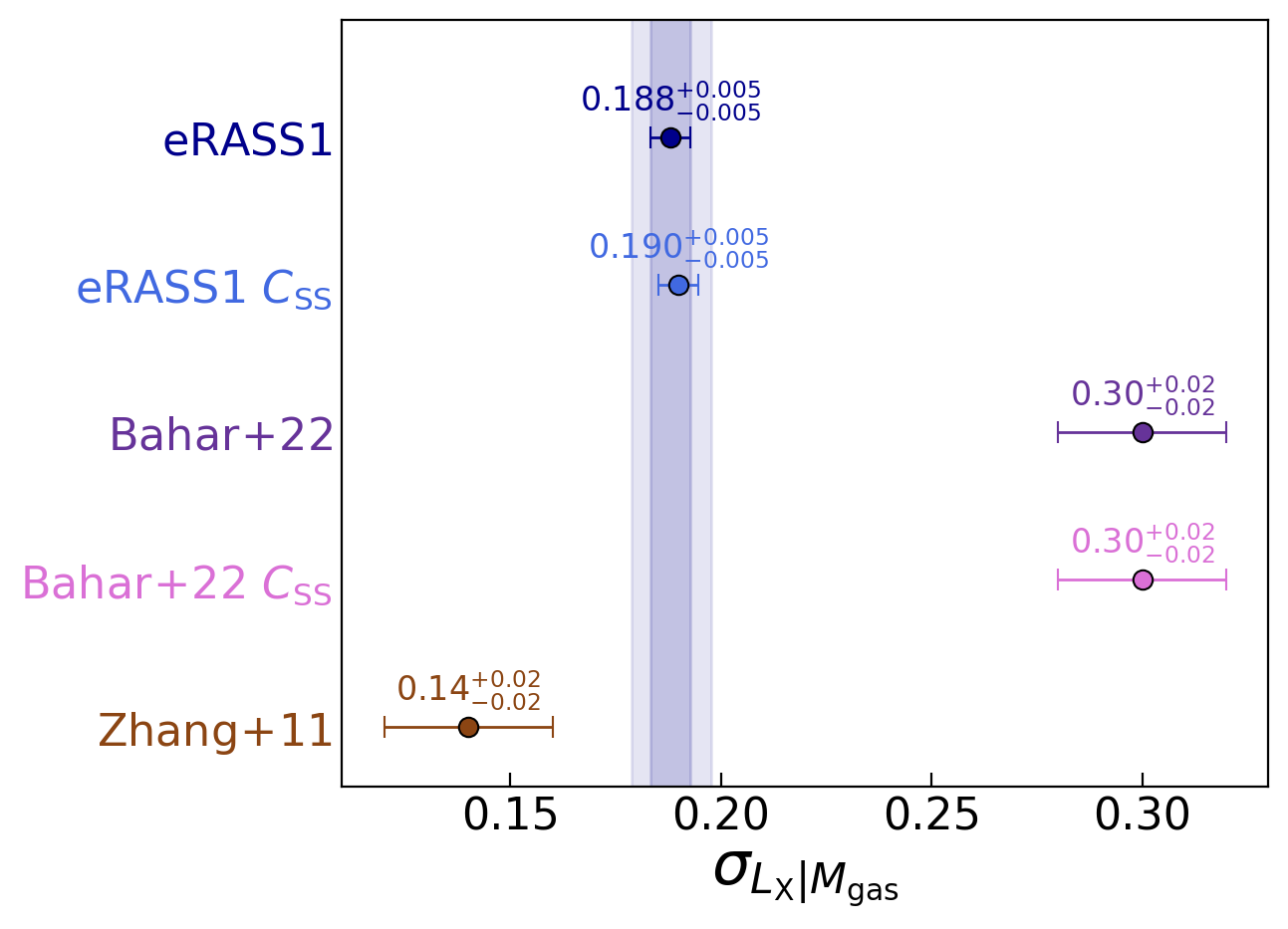}
    \caption{Same as Fig.~\ref{fig:Lx_T_sigma_comparison}, but presenting a comparison of the intrinsic scatter $\sigma_{L_{\rm X}|M_{\rm gas}}$ for the $L_{\rm X}-M_{\rm gas}$ scaling relation, including results from \citet[eFEDS cluster sample]{Bahar2022} and \cite{Zhang2011}.}
    \label{fig:Lx_Mgas_sigma_comparison}
\end{figure}

\begin{figure}
    \centering
    \includegraphics[width=0.8\linewidth]{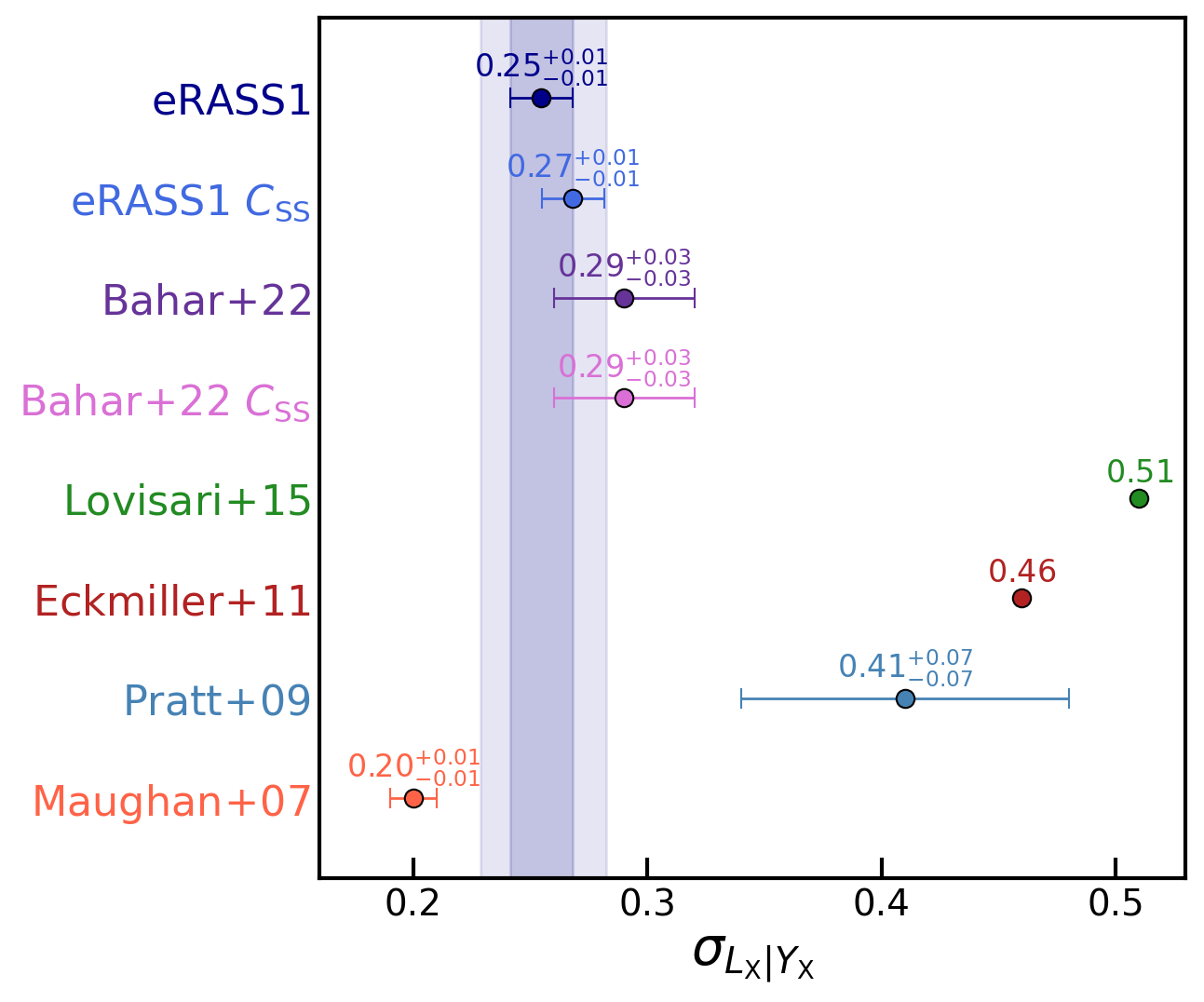}
    \caption{Same as Fig.~\ref{fig:Lx_T_sigma_comparison}, but showing a comparison of the intrinsic scatter $\sigma_{L_{\rm X}|Y_{\rm X}}$ for the $L_{\rm X}-Y_{\rm X}$ scaling relation, including results from \citet[eFEDS cluster sample]{Bahar2022}, \cite{Lovisari2015}, \cite{Eckmiller2011}, \cite{Pratt2009}, and \cite{Maughan2007}.}
    \label{fig:Lx_Yx_sigma_comparison}
\end{figure}

Figures~\ref{fig:Lx_T_sigma_comparison}, \ref{fig:Lx_Mgas_sigma_comparison}, and \ref{fig:Lx_Yx_sigma_comparison} display a comparison between our best-fit intrinsic scatter results with previous
studies, for the $L_{\rm X}-T$, $L_{\rm X}-M_{\rm gas}$ and $L_{\rm X}-Y_{\rm X}$ scaling relation analysis, respectively.

\begin{figure}
    \centering
    \includegraphics[width=0.8\linewidth]{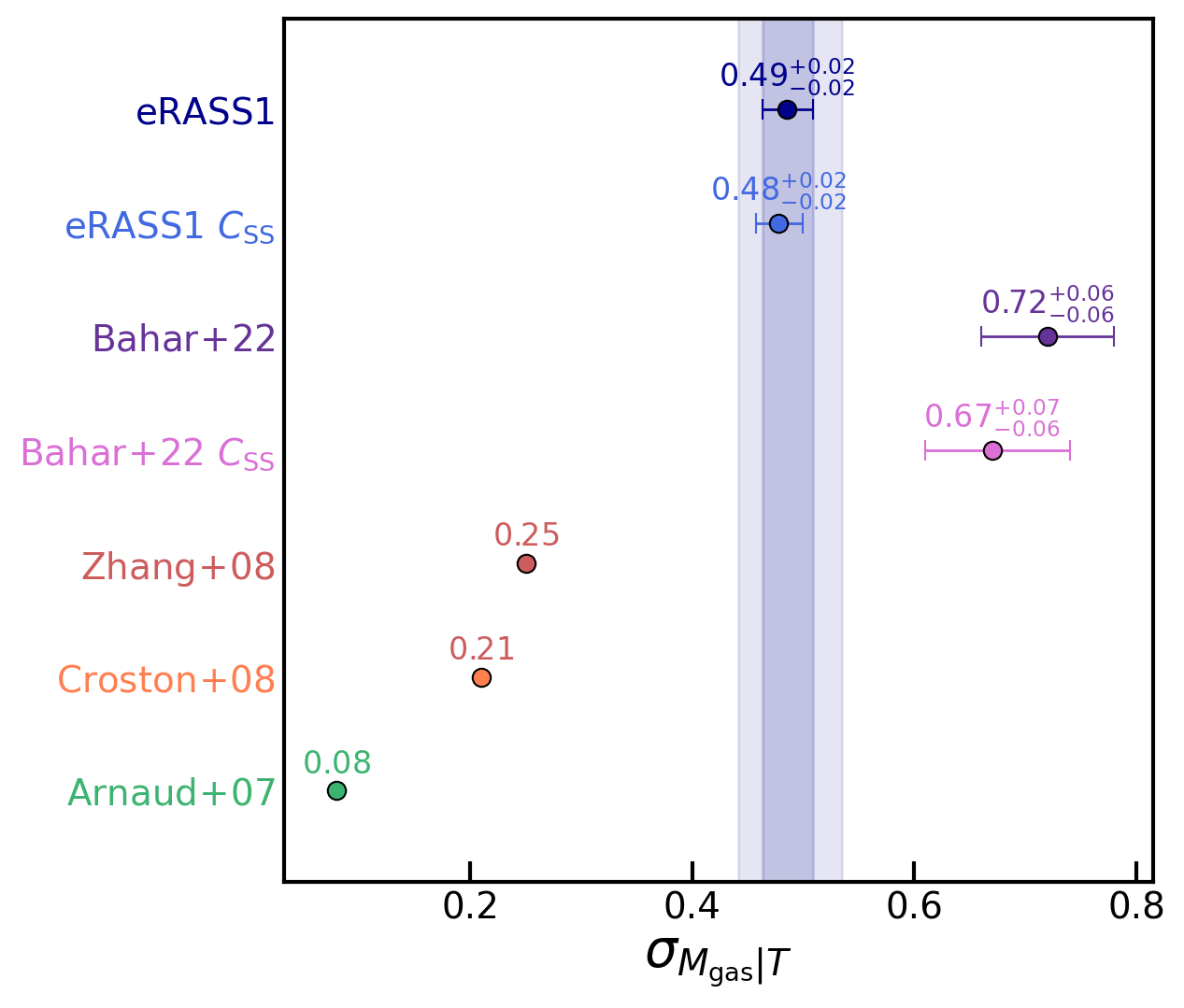}
    \caption{Same as Fig.~\ref{fig:Lx_T_sigma_comparison}, but showing a comparison of the intrinsic scatter $\sigma_{M_{\rm gas}|T}$ for the $M_{\rm gas}-T$ scaling relation, including results from \citet[eFEDS cluster sample]{Bahar2022}, \cite{Zhang2008}, \cite{Croston2008}, and \cite{Arnaud2007}.}
    \label{fig:Mgas_T_sigma_comparison}
\end{figure}

\end{appendix}

\end{document}